\documentclass[english]{elsarticle}
\usepackage[T1]{fontenc}
\usepackage[latin9]{inputenc}
\usepackage{geometry}
\usepackage{xcolor}
\usepackage{babel}
\usepackage{array}
\usepackage{units}
\usepackage{multirow}
\usepackage{amsthm}
\usepackage{amsmath}
\usepackage{graphicx}
\usepackage{setspace}
\PassOptionsToPackage{normalem}{ulem}
\usepackage{ulem}
\usepackage[unicode=true,
 bookmarks=true,bookmarksnumbered=false,bookmarksopen=false,
 breaklinks=false,pdfborder={0 0 1},backref=false,colorlinks=false]
 {hyperref}
\usepackage{breakurl}

\makeatletter

\providecommand{\tabularnewline}{\\}
\providecolor{lyxadded}{rgb}{0,0,1}
\providecolor{lyxdeleted}{rgb}{1,0,0}

\numberwithin{equation}{section}
\numberwithin{figure}{section}

\journal{Journal of Theoretical Biology}

\usepackage{savesym}
\savesymbol{eqref}
\usepackage[version=3]{mhchem}
\restoresymbol{dmb}{eqref}
\usepackage{lineno} 

\@ifundefined{showcaptionsetup}{}{%
 \PassOptionsToPackage{caption=false}{subfig}}
\usepackage{subfig}
\makeatother

\begin{document}

\begin{frontmatter}{}

\title{\textcolor{black}{The effects of MAPK activity on cell-cell adhesion
during wound healing}}

\author[APPM]{John T. Nardini\textcolor{black}{}}

\author[CH]{Douglas A. Chapnick\textcolor{black}{}}

\author[CH]{Xuedong Liu\textcolor{black}{}}

\author[APPM]{\textcolor{black}{David.$\ $M.$\ $Bortz}$^{*}$}

\ead{dmbortz@colorado.edu}

\address[APPM]{Department of Applied Mathematics, University of Colorado, Boulder,
CO, United States\textcolor{black}{{} 80309-0526 }}

\address[CH]{\textcolor{black}{Department of Chemistry and Biochemistry, University
of Colorado, Boulder, CO, United States 80303-0596.}}

\cortext[APPM]{Corresponding author D.M. Bortz\textcolor{black}{}}
\begin{abstract}
\begin{doublespace}
\textcolor{black}{The mechanisms underlying }\textcolor{black}{\emph{collective
migration}}\textcolor{black}{, or the coordinated movement of a population
of cells, are not well understood despite its ubiquitous nature. As
a means to investigate collective migration, we consider a wound healing
scenario in which a population of cells fills in the empty space left
from a scratch wound. Here we present a simplified mathematical model
that uses reaction-diffusion equations to model collective migration
during wound healing with an emphasis on cell movement and its response
to both cell signaling and cell-cell adhesion. We use the model to
investigate the effect of the MAPK signaling cascade on cell-cell
adhesion during wound healing after EGF treatment. Our results suggest
that activation of the MAPK signaling cascade stimulates collective
migration through increases in the pulling strength of leader cells.
We further use the model to suggest that treating a cell population
with EGF converts the time to wound closure (as function of wound
area) from parabolic to linear. }\end{doublespace}
\end{abstract}
\begin{keyword}
\texttt{Collective Migration \sep Cell signaling \sep Reaction-Diffusion
Equations \sep Wound healing \sep Epithelial-to-Mesenchymal Transition }
\end{keyword}

\end{frontmatter}{}

\begin{doublespace}

\section{\textcolor{black}{Introduction}}
\end{doublespace}

\begin{doublespace}
\textcolor{black}{Collective migration, or the coordinated movement
of a population of cells is a ubiquitous biological process involved
in tissue formation and repair, developmental regulation, and tumorigenesis
\citep{Rorth2009}. Although there has been a heavy focus on collective
migration in recent scientific literature, the mechanisms underlying
this important process are still not well understood \citep{Baker2012,Chapnick2014,Friedl2009,Nikolic2006}.
Determining the underlying mechanisms behind collective migration
behavior will likely provide insight into wound healing, as well as
the epithelial-to-mesenchymal transition (EMT) and metastasis characteristics
of tumor progression during cancer \citep{Chapnick2011}.}

\textcolor{black}{Significant intercellular communication is required
for a group of cells to coordinate their movement as one unit during
collective migration. Recent studies have sought to understand how
activation of the mitogen-activated protein kinase (MAPK) signaling
cascade may promote collective migration \citep{Chapnick2014,Huang2004}.
The MAPK signaling cascade affects many cellular activities, ranging
from cellular migration to angiogenesis \citep{Huang2004,Matsubayashi2004,Nikolic2006,Shin2001,Yang2013,Yewale2013}.
While several types of MAPKs appear to influence cell migration, the
extracellular signal-regulated kinas 1/2 (ERK) MAPKs are dominant
regulators of cellular migration. ERK influences cellular migration
in a variety of cell types and is downstream from the epidermal growth
factor receptor (EGFR) \citep{Jo2002,Klemke1997a,Webb2000}. EGFR
is a tyrosine kinase receptor that forms an active dimer state in
the presence of ligand, and its main ligand is epidermal growth factor
(EGF). Receptor dimerization initiates autophosphorylation of the
intracellular EGFR tyrosine kinase domains, which recruit signal transducers,
such as the Ras protein. These transducers then trigger various intracellular
signaling cascades, most notably the MAPK and ERK signaling cascades
\citep{Herbst2004,Yewale2013}. Accordingly, we focus on the activation
of ERK in response to EGF treatment in this study. }
\end{doublespace}

\textcolor{black}{Collective migration is defined by three main characteristics:
physical and functional connection between cells, multicellular polarity,
and the deposition and remodeling of the extracellular matrix \citep{Friedl2009}.
We focus on the first of these hallmarks through cell-cell adhesion.
We also note, however, that many aspects of }\textcolor{black}{\emph{single}}\textcolor{black}{{}
cell migration, such as molecular control of protrusions, polarity,
shape generation, and cell-surface adhesion also affect collective
migration \citep{DiMilla1991,Friedl2010,Holmes2012,Mori2008a,Palecek1997}.
Cell-cell adhesion is believed to have a stronger influence on collective
migration than these aforementioned single cell features, yet its
role is still poorly understood. \citep{Friedl2009,Friedl2010}. For
this reason, we ignore the effects of cell-surface adhesion and focus
on cell-cell adhesion, as has been done in previous studies on collective
migration \citep{Anguige2009,Johnston2012,Thompson2012,Tremel2009}. }

\textcolor{black}{Adherens junctions mediate cell-cell adhesion in
tissues by attaching actin and intermediate filament cytoskeletons
between adjacent cells, thereby allowing for dynamic coordinate force
transmission between the actin cytoskeletons of physically contacting
cells \citep{Friedl2009}. Adherens junctions form between cells through
transmembrane cadherin proteins, which bind actin filaments through
zyxin, vinculin, and $\alpha$-catenin adapter proteins \citep{Haley2009,Juliano2002}.
Cadherin proteins have been reported as both promoters and inhibitors
of cell migration, so their overall role is not clear \citep{Friedl2009,Geisbrecht2002,Hazan2000,Omelchenko2012}.
Based on this uncertainty in the role of cell-cell junctions during
collective migration and the MAPK cascade's well-established (stimulating)
role \citep{Chapnick2014,Huang2004,Matsubayashi2004,Nikolic2006},
we pose the following question: how does MAPK activity influence the
main role of cell-cell junctions during collective migration?}

\begin{doublespace}
\textcolor{black}{To provide insight into collective migration, we
will concentrate on wound healing induced in a sheet of cells from
the human keratinocyte (HaCaT) cell line. Experiments, as previously
described in \citep{Chapnick2014}, are conducted in which a wound
comprised of open space is formed in a monolayer of cells by scratching
away a portion of the population with a pipette tip. The remaining
cell population begins a collective migration process into the wound
and eventually fills in and closes the wound. Studies of this sort
tend to split cells in the population into two distinct classifications:
leader cells (those located at the wound interface) and follower cells
(those located farther away from the wound). We use live-cell imaging
to capture the cell profiles over time (every half hour) for 30 hours
and analyze both untreated (denoted as }\textcolor{black}{\emph{mock}}\textcolor{black}{)
and EGF-treated cell populations to analyze the effects of both EGF
and the MAPK signaling cascade on collective migration.}
\end{doublespace}

\textcolor{black}{Mathematical models can provide insight into how
EGF treatment stimulates collective migration, which may be difficult
to decipher through experimental procedures. Some authors strongly
advocate for mathematical models as necessary for a comprehensive
understanding of signaling cascades due to their complex quantitative
nature \citep{Clarke2008}. Accordingly, we focus on developing a
simple first mathematical model to investigate the relationship between
collective migration, the MAPK cascade, and cell-cell adhesion. In
particular, we will develop two models to investigate how the MAPK
signaling cascade affects cell-cell adhesion in response to EGF treatment
during wound healing. }

\textcolor{black}{Other studies have used various quantitative approaches
to model collective migration. A partial differential equation (PDE)-based
model is advantageous for modeling collective migration, as it allows
for an in-depth analysis of the biological system paired with the
ability to search for certain patterns, such as traveling wave solutions
to the model \citep{CurtisBortzFrontProp,HammondBortz2011}. This
class of wave solutions frequently arise in models of biological systems
\citep{Holmes2012,Langebrake2014a,Roques2012} as well as in collective
migration models due their ability to capture invasive wave front
dynamics}\footnote{\textcolor{black}{Discrete random walk models have also been proposed
to model collective migration and can yield similar mean-field results
to PDE models \citep{Baker2012,Plank2013}.}}\textcolor{black}{{} \citep{Maini2004TE,Murray2001}. Fisher's equation
(originally used to model the spread of advantageous genes \citep{Fisher1937})
is the classical example of a PDE model with traveling wave solutions,
and it has been used previously to model collective migration in response
to the wounding of a human peritoneal mesothelial cell population
\citep{Maini2004TE}.}

\textcolor{black}{A more recent study developed a two-dimensional
continuum model based on the force of lamellipodia, cell-substrate
adhesion, and cell-cell adhesion that can reproduce results from both
contraction experiments of enterocytes and expansion experiments of
Madin-Darby canine kidney (MDCK) cells \citep{Arciero2011}. A later
study demonstrated that simulations from this model also matched the
leading edge velocity of experimental data in intestinal enterocyte
(IEC) cells \citep{Arciero2013}. This model was further used as a
means to identify the effects of wound shape and area on the time
for wound closure. Ultimately, the authors concluded that the rate
of change of the wound area during wound healing is likely proportional
to the square root or first power of the wound area in IEC cells.
This was an improvement over previous studies that assumed the rate
of change of the wound area is constant, which yielded inaccurate
estimates for larger wounds \citep{Gilman2004}.}

\textcolor{black}{In \citep{Posta2010}, the authors computationally
recreated the results of \citep{Nikolic2006}, in which two waves
of MAPK activation were observed in MDCK cells during wound healing.
The authors of \citep{Posta2010} developed a four-species PDE model
for MAPK activation during wound healing that can quantitatively recapitulate
these two MAPK activation waves. The components of this model include
an arbitrary diffusible ligand (such as EGF), its cell-surface receptor,
an intracellular precursor to this ligand, and reactive oxygen species.
This study ignored cell migration, however, and thus did not investigate
the effect of these waves on wound healing. Other previous PDE models
have focused on how cell signaling effects cell migration. In \citep{Sherratt1990},
the authors showed how one simple regulatory chemical (impacting only
cell proliferation) is sufficient for a simple reaction-diffusion
model to match data on wound healing of epidermal cells. }

\textcolor{black}{In this study, we will develop two testable PDE
models using reaction-diffusion equations in order to elucidate properties
of the interaction between MAPK signaling, cellular migration, and
cell-cell adhesion in a wounded HaCaT cell population in response
to EGF treatment. These models will aid in making predictions on cellular
migration in epithelial cell populations.}

\textcolor{black}{We use these two models to test two competing hypotheses
regarding the influence of MAPK activation on the role of cell-cell
adhesion during wound healing. The first hypothesis states that }\textcolor{black}{\emph{MAPK
activity stimulates collective migration through decreases in the
drag strength of follower cells}}\textcolor{black}{. We denote the
corresponding mathematical model as Model H since cell-cell adhesion
will }\textcolor{black}{\emph{hinder}}\textcolor{black}{{} migration.
The second hypothesis states that }\textcolor{black}{\emph{MAPK activity
stimulates collective migration through increases in the pulling strength
of leader cells}}\textcolor{black}{. We denote the corresponding mathematical
model as Model P since cell-cell adhesion will }\textcolor{black}{\emph{promote}}\textcolor{black}{{}
migration. While both models are able to fit leading edge data, we
demonstrate that the model based on the second hypothesis can better
match various characteristics of the cell population during wound
healing. These model results suggest that the dominant role of cell-cell
adhesion in response to MAPK activation after EGF treatment is to
promote leader cell pulling rather than inhibit the drag of follower
cells.}

\begin{doublespace}
\textcolor{black}{In Section \ref{sec:Model-Development}, we present
our two model derivations based on the above hypotheses. In Section
\ref{sec:Results}, we demonstrate how Model P can match the leading
edge dynamics better than Model H and then use Model P to predict
wound closure as a function of wound area in HaCaT cell populations.
Lastly, in Section \ref{sec:Discussion}, we conclude with a discussion
on the implications of these results as well as plans for future work.}
\end{doublespace}

\begin{doublespace}

\section{\textcolor{black}{Model Development\label{sec:Model-Development}}}
\end{doublespace}

\textcolor{black}{Our mathematical models presented here consist of
two coupled variables describing cell density, $u(t,x)$, and the
average cellular MAPK activation level, $m(t)$, where our independent
variables include time ($t$) and spatial location ($x$). While we
are investigating cell-cell interactions in this study, current techniques
cannot directly measure these interactions. Hence, we use cell density
as a means to investigate cell-cell interactions during migration,
because it is measurable and indicative of physical interaction, so
high cell densities should display a larger degree of cell-cell interaction
effects on migration. We include the diffusion of cells and focus
on its response to different levels of MAPK activity. This can be
quantitatively described using a conservation law framework as:}

\textcolor{black}{
\begin{equation}
u_{t}(t,x)=\nabla\cdot(f_{u}^{D}\nabla u),\qquad m_{t}(t)=f_{m}^{A},\label{eq:outline_quant}
\end{equation}
where the subscript $t$ denotes differentiation with respect to time,
$f_{u}^{D}$ denotes the rate of cellular diffusion and $f_{m}^{A}$
describes the activation rate of $m(t)$. It should be noted that
cell proliferation and death could be included in Equation (\ref{eq:outline_quant}),
but are assumed negligible over the course of the experiment}\footnote{\textcolor{black}{The i}nclusion of a growth term had a minimal effect
on model simulations and hypothesis evaluation (results not presented
in this work).}\textcolor{black}{. Decreases in MAPK activation could also be considered
in Equation (\ref{eq:outline_quant}), however, we do not see any
noticeable decreases in MAPK levels during EGF experiments, so we
exclude them as well. Note that while we simulate our model in only
one spatial dimension due to the one-dimensional plane wave-behavior
in the wound healing data, the model can be easily extended to two
or three dimensions.}

We will develop two different models in Section \ref{sub:Models-for-diffusion}:
Model H focuses on how cell movement towards the wound may be hindered
by the drag of follower cells, while Model P focuses on how leader
cells may pull on cells behind them to promote movement towards the
wound. While we generally think of follower and leader cells in terms
of their location in the population, we assume that each cell behind
the first row in our population will either pull on those behind it
or hinder those in front of it. The resulting models thus distinguish
the dominant interaction (dragging or pulling) betweens neighboring
cells in the population during wound healing.

\begin{doublespace}
\textcolor{black}{In the rest of this section, we describe the development
of two simple models for different hypotheses on the emergence of
collective migration during wound healing. In Section \ref{sub:Model-1-for},
we describe our first diffusion term in which cell-cell adhesions
hinder motion, while in Section \ref{sub:Model-2-forEGF}, we describe
a different diffusion model where cell-cell adhesions promote motion.
Section \ref{sub:Cell-Growth} details our assumptions on MAPK activation
in response to EGF treatment. In Section \ref{sub:Models-for-EGF},
we then combine the models from Sections \ref{sub:Model-1-for} and
\ref{sub:Model-2-forEGF} with the observations on MAPK activation
in Section \ref{sub:Cell-Growth} to describe our models for EGF data
in Sections \ref{sub:Model-H-forEGF} and \ref{sub:Model-P-forEGF},
respectively. We then detail the entire model, including initial and
boundary conditions, in Section \ref{sub:Initial-and-boundary} and
outline our parameter estimation procedure in Section \ref{sub:Parameter-estimation}. }
\end{doublespace}

\subsection{\textcolor{black}{Models for diffusion\label{sub:Models-for-diffusion}}}

\textcolor{black}{In this section, we will derive two different models
(Models H and P) for diffusion based off cell-cell adhesion. In order
to do so, we first discretize our solution domain. We use a uniform
grid for both the time and space domain, so the time interval may
be denoted as $t(j)=j\Delta t,\ j=0,...,N-1$, where $N$ denotes
the total number of time points used, and $x(i)=i\Delta x,i=0,...,M-1$
where $M$ denotes the number of spatial points used. We simplify
notation by writing $t(j)=t_{j},$ $x(i)=x_{i},$ and write our discretized
solution as $u(t_{j},x_{i})=u_{i}^{j}$. }

\subsubsection{\textcolor{black}{Model H: cell-cell adhesions hinder migration\label{sub:Model-1-for}}}

\begin{doublespace}
\textcolor{black}{For our first model on the effects of cell-cell
adhesion on diffusion, we assume that cell-cell adhesions }\textcolor{black}{\emph{hinder}}\textcolor{black}{{}
migration through the drag of follower cells and accordingly denote
it as Model H. In this scenario, the downregulation of cell-cell adhesions
will promote migration. We create this model by modifying a previous
cell migration study that incorporates cell-cell adhesion.}

\textcolor{black}{In \citep{Anguige2009}, the authors developed a
mathematical model that assumes cell-cell adhesions hinder cell movement.
To derive the model, a probability transition $\tau_{i}^{+}$ is given
for a cell density located at position $x_{i}$ attempting to move
forward to location $x_{i+1}$ by}

\textcolor{black}{
\begin{equation}
\tau_{i}^{+}=\dfrac{(1-u_{i+1})(1-\mbox{\ensuremath{\alpha}}u_{i-1})}{\Delta x^{2}}\label{eq:transition_prob}
\end{equation}
where the first factor represents space filling (e.g., the cell density
is more likely to move forward given a smaller cell density in front
of it). The parameter $\alpha$ denotes the (dimensionless) rate of
cell-cell adhesion, and thus }\textcolor{black}{\emph{the second term
in Equation (\ref{eq:transition_prob}) represents our first hypothesis
in which follower cells (denoted with $u_{i-1}$) hinder the movement
of cells in front of them towards the wound.}}\textcolor{black}{{} The
transition probabilities $\tau_{i-1}^{+},\tau_{i}^{-},\text{ and }\tau_{i+1}^{-}$
are all defined analogously (with the minus signs denoting cells moving
backwards). We note that Equation (\ref{eq:transition_prob}) may
be modified to include the influence of other aspects of cell migration,
including cell-surface adhesion and shape change \citep{Friedl2010}. }

\textcolor{black}{Given these transition probabilities, Anguige and
Schmeiser derive the continuum limit of the model by analyzing the
change of cell density $u_{i}$ over time as
\begin{equation}
u_{t}=\tau_{i-1}^{+}u_{i-1}+\tau_{i+1}^{-}u_{i+1}-(\tau_{i}^{+}+\tau_{i}^{-})u_{i}\label{eq:disc_model1}
\end{equation}
and then taking the limit as $\Delta x\rightarrow0$. The derivation
of the continuum limit of Equation (\ref{eq:disc_model1}) is presented
in \citep{Anguige2009} and may be written as the dimensionless model
\[
u_{t}=((1+3\alpha(u-\nicefrac{2}{3})^{2}-\nicefrac{4}{3}\alpha)u_{x})_{x},
\]
and we accordingly write the dimensional model as
\begin{equation}
u_{t}=((D+3\gamma(u-\nicefrac{2}{3})^{2}-\nicefrac{4}{3}\gamma)u_{x})_{x}.\quad\mbox{(Model H)}\label{eq:f_modelh}
\end{equation}
In Equation (\ref{eq:f_modelh}), $D$ and $\gamma$ denote the rates
of diffusion and cell-cell adhesion with units $\nicefrac{\mbox{microns}^{2}}{hr}$.
Hence the diffusion rate for Model H is
\[
f_{u}^{D}=D+3\gamma(u-\nicefrac{2}{3})^{2}-\nicefrac{4}{3}\gamma.
\]
We note that backwards diffusion may occur with Equation (\ref{eq:f_modelh})
if $4\gamma>3D$ and address this inequality more after final model
development in Section \ref{sub:Model-H-forEGF}.}
\end{doublespace}

\subsubsection{\textcolor{black}{Model P: cell-cell adhesions promote}\textcolor{black}{\emph{
}}\textcolor{black}{migration \label{sub:Model-2-forEGF}}}

\begin{doublespace}
\textcolor{black}{For our second model, we assume that cell-cell adhesions
}\textcolor{black}{\emph{promote}}\textcolor{black}{{} migration as
leader cells pull the cells behind them forward and denote it as Model
P. In this scenario, the upregulation of cell-cell adhesion will promote
migration. We now denote the forward transition probability $\tau_{i}^{+}$
of cell density $u_{i}$ as
\begin{equation}
\tau_{i}^{+}=\dfrac{(1-u_{i+1})(1+\mbox{\ensuremath{\alpha}}u_{i+1})}{\Delta x^{2}},\label{eq:trans_prob_2}
\end{equation}
where the first term again represent space filling, but}\textcolor{black}{\emph{
the second term in Equation (\ref{eq:trans_prob_2}) represents our
second assumption in which leader cells (denoted with $u_{i+1})$
promote movement towards the wound by pulling on the cells behind
them.}}\textcolor{black}{{} We note again that $\tau_{i-1}^{+},\tau_{i}^{-},\text{ and }\tau_{i+1}^{-}$
are all defined analogously. Substituting the transition probability
from Equation (\ref{eq:trans_prob_2}) into Equation (\ref{eq:disc_model1})
yields }

\textcolor{black}{
\begin{eqnarray}
u_{t} & = & \dfrac{(1-u_{i})(1+\mbox{\ensuremath{\alpha}}u_{i})u_{i-1}+(1-u_{i})(1+\mbox{\ensuremath{\alpha}}u_{i})u_{i+1}-((1-u_{i+1})(1+\mbox{\ensuremath{\alpha}}u_{i+1})+(1-u_{i-1})(1+\mbox{\ensuremath{\alpha}}u_{i-1}))u_{i}}{\Delta x^{2}}\nonumber \\
 & = & \dfrac{u_{i-1}+\mbox{\ensuremath{\alpha}}u_{i-1}^{2}u_{i}-\mbox{\ensuremath{\alpha}}u_{i-1}u_{i}^{2}+u_{i+1}-2u_{i}-\mbox{\ensuremath{\alpha}}u_{i}^{2}u_{i+1}+\mbox{\ensuremath{\alpha}}u_{i}u_{i+1}^{2}}{\Delta x^{2}}\nonumber \\
 & = & \dfrac{u_{i-1}-2u_{i}+u_{i+1}}{\Delta x^{2}}+\mbox{\ensuremath{\alpha}}\dfrac{u_{i-1}^{2}u_{i}-u_{i-1}u_{i}^{2}-u_{i}^{2}u_{i+1}+u_{i}u_{i+1}^{2}}{\Delta x^{2}},\label{eq:model_2_deriv}
\end{eqnarray}
where we can recognize the first term on the right hand side as the
standard central difference approximation to the second derivative.
For the second term on the right hand side, we can simplify using
Taylor series approximations as: $u_{i+1}\approx u_{i}+\Delta x\cdot u_{i}'+\nicefrac{\Delta x^{2}}{2}\cdot u_{i}'',\;u_{i-1}\approx u_{i}-\Delta x\cdot u_{i}'+\nicefrac{\Delta x^{2}}{2}\cdot u_{i}''$,
where primes denote a spatial derivative term. Substituting these
terms into the second term on the right hand side of Equation (\ref{eq:model_2_deriv})
reveals
\begin{eqnarray*}
u_{t} & = & \mbox{\ensuremath{\dfrac{u_{i-1}-2u_{i}+u_{i+1}}{\Delta x^{2}}}+\ensuremath{\alpha}}\dfrac{2\Delta x^{2}u_{i}(u_{i}')^{2}+u_{i}^{2}u_{i}''\Delta x^{2}+\mathcal{O}(\Delta x^{4})}{\Delta x^{2}}\\
 & = & u_{i}''+\alpha(2u_{i}(u_{i}')^{2}+u_{i}^{2}u_{i}'')+\mathcal{O}(\Delta x^{2})\\
 & = & u_{i}''+\alpha(u_{i}^{2}u_{i}')'+\mathcal{O}(\Delta x^{2})
\end{eqnarray*}
resulting in the dimensionless continuum limit to be}
\end{doublespace}

\textcolor{black}{
\[
u_{t}=((1+\alpha u^{2})u_{x})_{x},
\]
which we dimensionalize as
\begin{equation}
u_{t}=((D+\gamma u^{2})u_{x})_{x},\quad\mbox{(Model P)}\label{eq:cont_lim_model2}
\end{equation}
where $D$ and $\gamma$ again denote the dimensionalized rates of
cell diffusion and cell-cell adhesion, respectively. Thus the rate
of diffusion for Model P is given as 
\[
f_{u}^{D}=D+\gamma u^{2}.
\]
We note that a more general form of this equation has been used previously
to model temperature fronts and can have traveling wave solutions
under certain scenarios \citep{samarskii1963}. }

\subsection{\textcolor{black}{MAPK activation and cell growth in response to
EGF treatment\label{sub:Cell-Growth}}}

\begin{figure}
\centering{}\textcolor{black}{\includegraphics[width=0.45\textwidth]{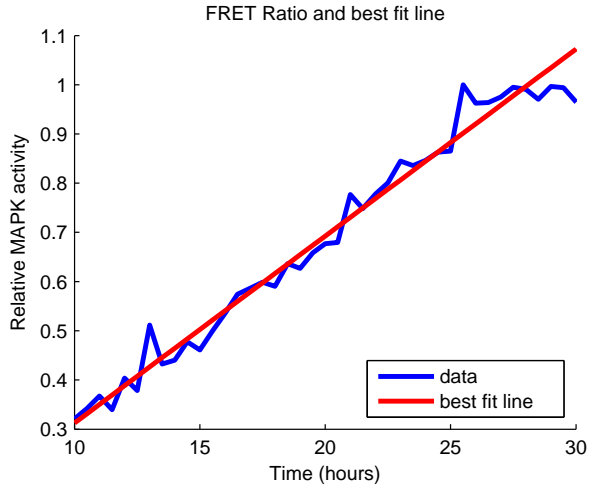}\protect\caption{Plotting the relative average FRET ratio of the cell population during
the EGF experiment from $t=10$ to $30$ hours, along with its best
fit line, which is given by $b(t)$ in the text. Note that this data
has an $R^{2}$ value of 0.98.\label{fig:FRET}}
}
\end{figure}

\textcolor{black}{While we have difficulty matching experimental data
over the entire experiment, our models performed well in the time
range of 10 hours after EGF treatment until the end of the experiment
at $t=30$ hours. In Figure \ref{fig:FRET}, we have plotted the normalized
average level of FRET ratio (indicative of MAPK activation in the
population) during the experimental protocol. The activation levels
after $t=10$ appear linear, so we focus on matching data in this
time interval ($t=10-30$ hours) and use a linear term for MAPK activation.
We have plotted a best-fit line to the normalized FRET ratio data
from $t=10-30$ hours in Figure \ref{fig:FRET}, which is given by
$b(t)=-0.0675+0.0380t$. Note that this data has an $R^{2}$ value
of 0.98 to help justify this linear term. We thus set $f_{m}^{A}$
as a constant, $c$, in Equation (\ref{eq:outline_quant}). Also note
that $b(t)>1$ after $t\approx28$ hours, even though the data has
been normalized, so we set $m(t)=1$ for these values. }

\textcolor{black}{}

\subsection{\textcolor{black}{Models for EGF data\label{sub:Models-for-EGF}}}

\textcolor{black}{We now present the final models for collective migration
in response to EGF treatment. We do so by assuming that the rate of
cell-cell adhesion changes over time in response to the level of $m$
activation. This requires a variable term, denoted by $\Gamma(m)$,
instead of a constant value, $\gamma$, for the rate of cell-cell
adhesion.}

\subsubsection{\textcolor{black}{Model H: MAPK stimulates collective migration through
decreases in drag strength\label{sub:Model-H-forEGF}}}

\begin{doublespace}
\textcolor{black}{To conclude Model H, we now extend $f_{u}^{D}$
in Equation (\ref{eq:f_modelh}) to include changes in cell-cell adhesion
in response to different levels of MAPK activation. This model is
consistent with the hypothesis that }\textcolor{black}{\emph{MAPK
activity stimulates collective migration through decreases in the
drag strength of follower cells}}\textcolor{black}{. We use a linear
cell-cell adhesion term as a function of MAPK activity, $m$, by $\Gamma(m)=\gamma(1-m)$,
for some cell-cell adhesion constant $\gamma>0$. Note that $0\leq m\leq1$
so that $\Gamma(m)\ge0$. }

\textcolor{black}{Accordingly, our final description for Model H in
response to EGF treatment is given by
\begin{eqnarray}
u_{t}(t,x) & =((D+3\Gamma(m)(u-\nicefrac{2}{3})^{2}-\nicefrac{4}{3}\Gamma(m))u_{x})_{x}, & m_{t}(t)=c\quad\mbox{(Model H for EGF)}\label{eq:model1}
\end{eqnarray}
where $\Gamma(m)$ is given above. We note that if $4D<3\Gamma(m)$,
the above equation will exhibit backwards diffusion, which is not
well-defined. Fortunately, all parameter sets in this study do not
satisfy this inequality for the entirety of the simulation.}
\end{doublespace}

\subsubsection{\textcolor{black}{Model P: MAPK activation stimulates collective
migration by increases in pulling strength\label{sub:Model-P-forEGF}}}

\begin{doublespace}
\textcolor{black}{We extend $f_{u}^{D}$ from Equation (\ref{eq:cont_lim_model2})
to include changes in cell-cell adhesion in response to different
levels of MAPK activation for Model P in response to EGF data. In
this scenario, we assume that }\textcolor{black}{\emph{MAPK activity
stimulates collective migration through increases in the pulling strength
of leader cells}}\textcolor{black}{. Accordingly, we model the cell-cell
adhesion level as a function of $m$ by $\Gamma(m)=\gamma m$, for
some constant $\gamma$. As a result, Model P for EGF-treated cells
is given by
\begin{eqnarray}
u_{t}(t,x) & =((D+\Gamma(m)u^{2})u_{x})_{x}+ku(1-u), & m_{t}(t)=c,\quad\mbox{(Model P for EGF)}\label{eq:model2_final}
\end{eqnarray}
where $\Gamma(m)$ is given above.}
\end{doublespace}

\begin{doublespace}

\subsection{\textcolor{black}{Complete models, initial and boundary conditions\label{sub:Initial-and-boundary}}}
\end{doublespace}

\begin{doublespace}
\textcolor{black}{We've now introduced the two models under consideration
in this study. We briefly review the two models, along with their
terms and assumptions in Table \ref{tab:models}. A list of all parameters
used throughout this study is also given in Table \ref{tab:List-of-parameters.}.}

\textcolor{black}{}
\begin{table}[h]
\centering{}\textcolor{black}{}%
\begin{tabular}{|c|c|c|c|c|}
\hline 
\textbf{\textcolor{black}{Model features}} & \multicolumn{2}{c|}{\textbf{\textcolor{black}{H}}} & \multicolumn{2}{c|}{\textbf{\textcolor{black}{P}}}\tabularnewline
\hline 
\textcolor{black}{Cell-cell adhesions} & \multicolumn{2}{c|}{\textcolor{black}{Hinder migration}} & \multicolumn{2}{c|}{\textcolor{black}{Promote migration}}\tabularnewline
\hline 
\textcolor{black}{Simulation} & \textcolor{black}{mock} & \textcolor{black}{EGF} & \textcolor{black}{mock} & \textcolor{black}{EGF}\tabularnewline
\hline 
\textcolor{black}{Equation} & \textcolor{black}{(\ref{eq:f_modelh})} & \textcolor{black}{(\ref{eq:model1})} & \textcolor{black}{(\ref{eq:cont_lim_model2})} & \textcolor{black}{(\ref{eq:model2_final})}\tabularnewline
\hline 
\textcolor{black}{MAPK impact on} & \multirow{2}{*}{\textcolor{black}{n/a}} & \multirow{2}{*}{\textcolor{black}{Downregulation}} & \multirow{2}{*}{\textcolor{black}{n/a}} & \multirow{2}{*}{\textcolor{black}{Upregulation}}\tabularnewline
\textcolor{black}{cell-cell adhesions} &  &  &  & \tabularnewline
\hline 
\textcolor{black}{$\Gamma(m)$} & \textcolor{black}{$\gamma$} & \textcolor{black}{$\gamma(1-m)$} & \textcolor{black}{$\gamma$} & \textcolor{black}{$\gamma m$}\tabularnewline
\hline 
\end{tabular}\textcolor{black}{\protect\caption{Summary of equations and assumptions used for Models H and P. \label{tab:models}}
}
\end{table}

\textcolor{black}{}
\begin{table}[h]
\centering{}\textcolor{black}{}%
\begin{tabular}{|c|c|c|c|c|c|c|}
\hline 
\multirow{2}{*}{\textbf{\textcolor{black}{Parameter}}} & \textbf{\textcolor{black}{Description }} & \multicolumn{2}{c|}{\textbf{\textcolor{black}{Model H}}} & \multicolumn{2}{c|}{\textbf{\textcolor{black}{Model P}}} & \textbf{\textcolor{black}{Determined}}\tabularnewline
\cline{3-6} 
 & \textbf{\textcolor{black}{(units)}} & \textbf{\textcolor{black}{Mock}} & \textbf{\textcolor{black}{EGF}} & \textbf{\textcolor{black}{Mock}} & \textbf{\textcolor{black}{EGF}} & \textbf{\textcolor{black}{by}}\tabularnewline
\hline 
\multirow{2}{*}{\textcolor{black}{$D$}} & \textcolor{black}{Baseline rate of diffusion } & \multirow{2}{*}{\textcolor{black}{19,770}} & \multirow{2}{*}{177,940} & \multirow{2}{*}{\textcolor{black}{2.19}} & \multirow{2}{*}{0.14} & \multirow{2}{*}{\textcolor{black}{Fitting}}\tabularnewline
 & \textcolor{black}{($\nicefrac{\mbox{microns}^{2}}{\mbox{hr}})$} &  &  &  &  & \tabularnewline
\hline 
\multirow{2}{*}{\textcolor{black}{$c$}} & \textcolor{black}{Rate of MAPK activation } & \multirow{2}{*}{\textcolor{black}{0}} & \multirow{2}{*}{\textcolor{black}{0.038}} & \multirow{2}{*}{\textcolor{black}{0}} & \multirow{2}{*}{\textcolor{black}{.038}} & \multirow{2}{*}{\textcolor{black}{$b(t)$}}\tabularnewline
 & \textcolor{black}{($\mbox{hr}^{-1}$)} &  &  &  &  & \tabularnewline
\hline 
\multirow{3}{*}{\textcolor{black}{$\gamma$}} & \textcolor{black}{Rate of adhesion between adjacent cells } & \multirow{3}{*}{\textcolor{black}{1.68}} & \multirow{3}{*}{\textcolor{black}{970}} & \multirow{3}{*}{\textcolor{black}{30,614}} & \multirow{3}{*}{535,810} & \multirow{3}{*}{\textcolor{black}{Fitting}}\tabularnewline
 & \textcolor{black}{($\nicefrac{\mbox{microns}^{2}}{\mbox{hr}}$) for
mock} &  &  &  &  & \tabularnewline
 & \textcolor{black}{($\nicefrac{\mbox{microns}^{2}}{\mbox{hr}\cdot m}$)
for EGF} &  &  &  &  & \tabularnewline
\hline 
\multirow{2}{*}{\textcolor{black}{$w$}} & \textcolor{black}{Location of wound } & \multirow{2}{*}{\textcolor{black}{1550}} & \multirow{2}{*}{\textcolor{black}{1620}} & \multirow{2}{*}{\textcolor{black}{1550}} & \multirow{2}{*}{\textcolor{black}{1620}} & \multirow{2}{*}{\textcolor{black}{From data}}\tabularnewline
 & \textcolor{black}{(microns)} &  &  &  &  & \tabularnewline
\hline 
\end{tabular}\textcolor{black}{\protect\caption{List of parameters for the models, along with the parameter values
used and how we obtained these values. \label{tab:List-of-parameters.}}
}
\end{table}

\begin{figure}
\textcolor{black}{}\subfloat[Experimental snap shots of the experiment and resulting profiles for
the mock experiments. Top row: Images of the wound healing experiment
at $t=7.5$ and 20 hours. Bottom row: Resulting profiles for the cell
population in the image\textcolor{blue}{.} ]{\textcolor{black}{\protect\includegraphics[width=0.45\textwidth]{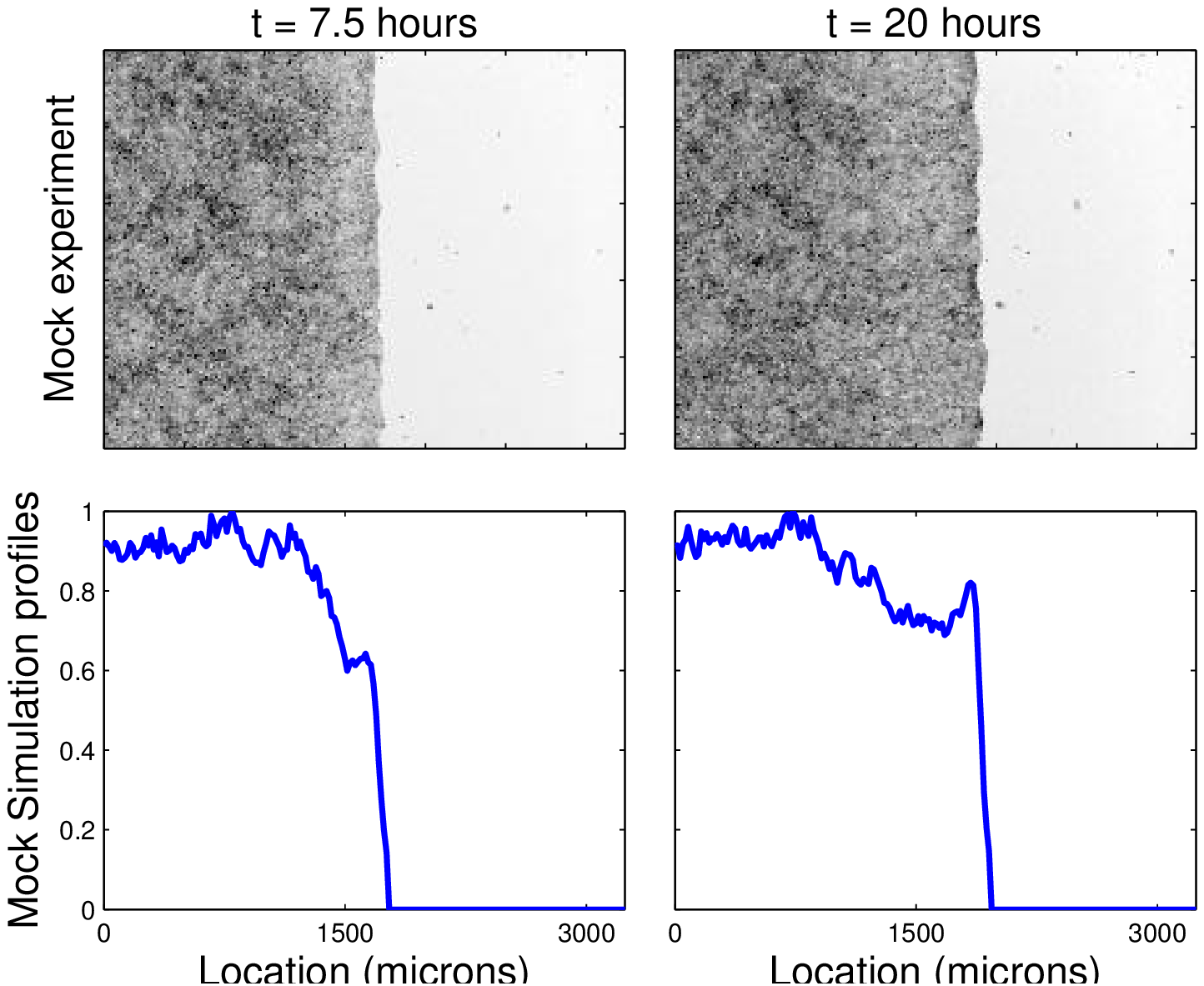}}}\textcolor{black}{\hfill{}}\subfloat[Experimental snap shots of the experiment and resulting profiles for
the EGF experiments. Top row: Images of the wound healing experiment
at $t=7.5$ and 20 hours. Bottom row: Resulting profiles for the cell
population in the image\textcolor{blue}{.} ]{\textcolor{black}{\protect\includegraphics[width=0.45\textwidth]{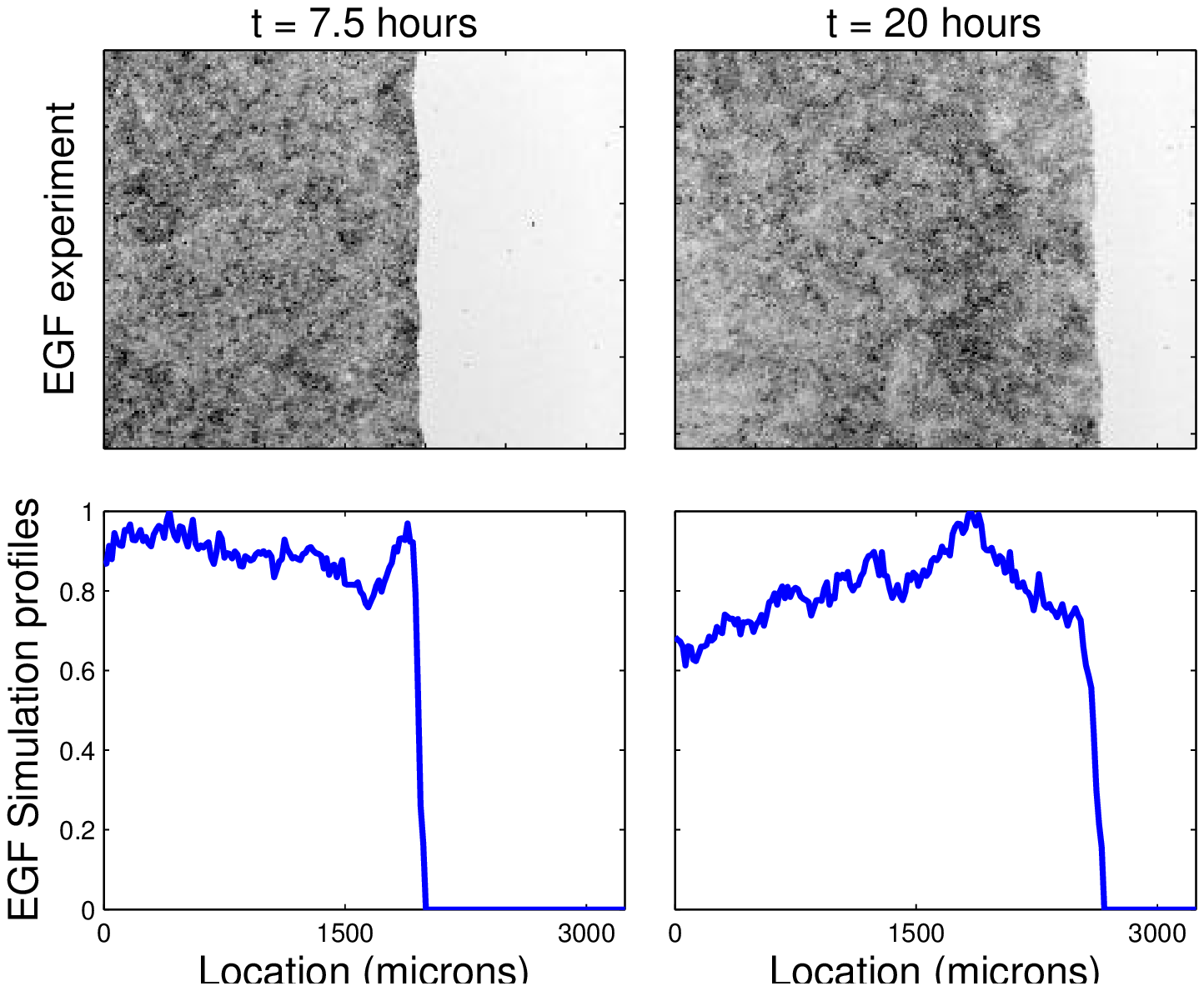}}

}

\textcolor{black}{}\subfloat[\textcolor{black}{Locations of the mock (blue) and EGF (red) experimental
leading edges over time using Equation (\ref{eq:LE_calc}), which
identifies the first point where $u(t_{j},x_{i})\le0.3$ for each
time point $t_{j}$, as is described in Section \ref{sub:Parameter-estimation}.
Note that the mock leading edge data may take the same value for several
consecutive time points because it moves slowly, and the camera discretizes
the experimental domain.} ]{\textcolor{black}{\protect\includegraphics[width=0.45\textwidth]{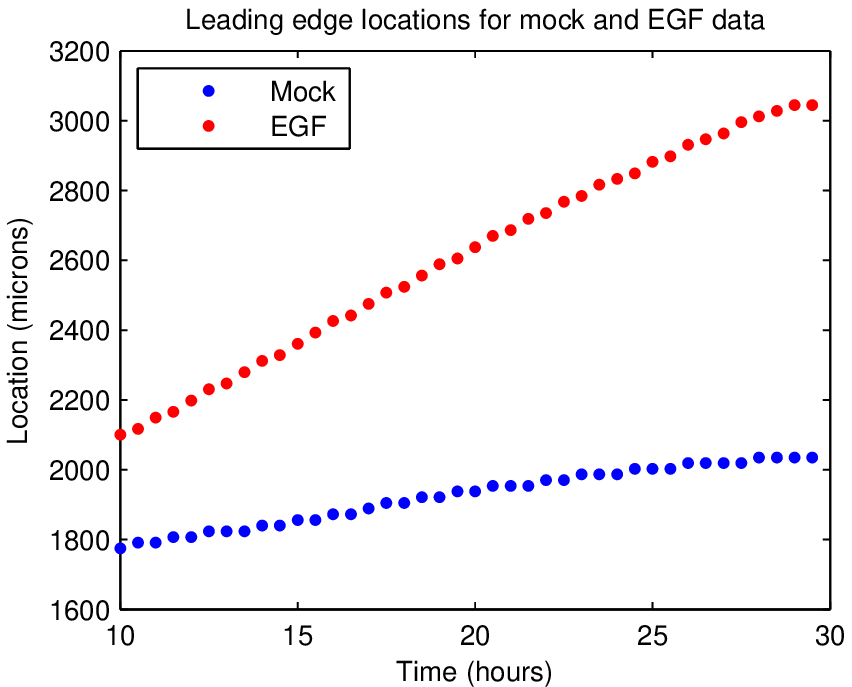}}}\textcolor{black}{\protect\caption{\label{fig:mock_experiment_IC}}
}
\end{figure}

\textcolor{black}{If we let $w$ denote the location of the wound,
then we would ideally use an initial condition that is nonzero for
all $x\le w$ to imitate the presence of cells and zero for all $x>w$
to imitate the wound. Scratching the cell population with the pipette
tip, however, initially shocks the cells and causes slow initial movement.
For this reason, we use a nonzero initial condition for all $x\le\tilde{w}=w-180$
microns and a zero initial condition for all $x>\tilde{w}$. This
alteration to the initial condition allows us to match the wave phenomena
of both the model and data. We compute the nonzero portion of the
initial condition by interpolating the data profile at the initial
time point. Thus, our initial condition is given by 
\begin{eqnarray}
u(t_{0},x) & = & \begin{cases}
\psi(x); & x\leq\tilde{w}\\
0; & \mbox{otherwise}
\end{cases}\nonumber \\
m(t_{0}) & = & b(t_{0}).\label{eq:IC}
\end{eqnarray}
where $t_{0}$ is our starting time for simulations (e.g., $t_{0}=10$
hours) and $\psi$ denotes the interpolated, normalized initial data
profile values at time $t_{0}$. In Figure \ref{fig:mock_experiment_IC},
we've included some typical video snapshots at $t=7.5,20$ hours,
along with their resulting density profile. In order to obtain this
population profile from data, we sum over the vertical axis for each
image matrix and then normalize over each time step. The linear interpolation
of this initial profile (at $t=t_{0})$ is our $\psi(x)$ function
used for the initial condition. Because of the linear MAPK activation
phase between $t=10-30$ hours, we will simulate the model from $t=7.5$
to 30 hours, where we also run the model for an extra 2.5 hours due
to its initial high speeds.}\textcolor{blue}{{} }\textcolor{black}{We
also include the leading edge locations of the two experiments in
the bottom portion of Figure \ref{fig:mock_experiment_IC}, whose
calculation is described in the following section. }
\end{doublespace}

\textcolor{black}{}

\begin{doublespace}
\textcolor{black}{We use zero Neumann boundary conditions to simulate
no flux conditions at the walls of the well plate. Note that in experimental
videos, we only observe a $3.24\ mm\times3.24\ mm$ field of view,
whereas the experimental domain is actually $7\ mm$ long and $5\ mm$
wide, so there is plenty of space for the cells to move in either
direction. Accordingly, we extend our computational domain 1.62 $mm$
behind the field of view and 3.24 $mm$ in front of the field of view
to ensure the boundaries do not affect the model simulations.}
\end{doublespace}

\subsection{\textcolor{black}{Parameter estimation\label{sub:Parameter-estimation}}}

\begin{doublespace}
\textcolor{black}{In order to fit our model simulations to data on
wound healing, we will use an inverse problem procedure to estimate
the parameter vector $\vec{q}=[D,\alpha]^{T}$ for both models for
both mock and EGF data. We do so by comparing the locations of the
leading edge for both the experimental data and model simulations,
as has been done in previous studies on wound healing \citep{Johnston2014,Maini2004TE}.
In this work, we will define the leading edge as the location in which
the relative density is equal to a certain value, although we note
that it has different definitions elsewhere \citep{Chapnick2014}.
As seen in Figures \ref{fig:mock_exp} and \ref{figLegf_exp}, a density
of 0.3 can locate the leading edge well. Accordingly, we define $\ell_{m,0.3}(t,\mbox{\ensuremath{\hat{q}}})$
as the leading edge location over time for the model using $\hat{q}$
as an estimate for the true $\mbox{\ensuremath{\vec{q}}}$, denoted
by
\[
\ell_{m,0.3}(t,\mbox{\ensuremath{\hat{q}}})=\left\{ x\left|u(x,t)=0.3\right.,\mbox{\ensuremath{\vec{q}}=}\mbox{\ensuremath{\hat{q}}}\right\} .
\]
Note that with numerical computations, we cannot precisely estimate
the leading edge in this manner. Accordingly, we estimate the leading
edge to be the first location where the relative density is below
the set threshold. For instance, if $\ell_{m,0.3}^{n}(\mbox{\ensuremath{\hat{q}}})$
denotes the leading edge location at time $t_{n}$, then we approximate
it as:}

\textcolor{black}{
\begin{equation}
\ell_{m,0.3}^{n}(\mbox{\ensuremath{\hat{q}})}\approx\hat{\ell}_{m,0.3}^{n}(\mbox{\ensuremath{\hat{q}})}=\left\{ x_{i}\left|u_{i}^{n}\leq0.3\text{ and }u_{k}^{n}>0.3\ \mbox{for all}\ k<i\right.,\mbox{\ensuremath{\vec{q}}=\ensuremath{\hat{q}}}\right\} .\label{eq:LE_calc}
\end{equation}
We use analogous definition for the estimate of the leading edge of
the normalized data over time, which is denoted as $\hat{\ell}_{D,0.3}^{n}$.}

\textcolor{black}{As a means to estimate $\vec{q}$ using our models,
we will implement an inverse problem in which we minimize a cost function
\citep{BanksTran2009}. In this work, we will use the cost function
given by an ordinary least squares estimate: 
\[
J(\vec{q})=\sum_{n=1}^{N}|\hat{\ell}_{m,0.3}^{n}(\vec{q})-\hat{\ell}_{D,0.3}^{n}|^{2},
\]
where $N$ is the number of time points considered. To find the value
of $\vec{q}$ that minimizes the cost function, we use the Nelder-Mead
algorithm as implemented in MATLAB's fminsearch command.}
\end{doublespace}

\begin{doublespace}

\section{\textcolor{black}{Results\label{sec:Results}}}
\end{doublespace}

\begin{doublespace}
\textcolor{black}{To investigate the performance of Models H and P
in describing the wound healing process, we now compare both models
to mock and EGF experimental data. To compare these prediction to
the experimentally determined results, we use images of the cell populations
over time, such as those depicted in Figure \ref{fig:mock_experiment_IC}.
We use several different criteria for comparison to the experimental
data. In Section \ref{sub:Leading-edge-fitting}, we compare the leading
edge locations of the model simulations and experimental data for
both of the mock and EGF cases. In Section \ref{sub:Wavespeed-comparisons},
we compare the wave speed values of the EGF model simulations to the
experimental data wave speed values. In Section \ref{sub:Comparison-to-data},
we directly compare the experimental snapshots to the model profiles
over time. We conclude the results section by using Model P to investigate
the time to wound closure as a function of the wound area in Section
\ref{sub:Time-to-wound}. }
\end{doublespace}

\subsection{\textcolor{black}{Leading edge fitting\label{sub:Leading-edge-fitting}}}

\begin{figure}
\textcolor{black}{\centering\includegraphics[width=0.45\textwidth]{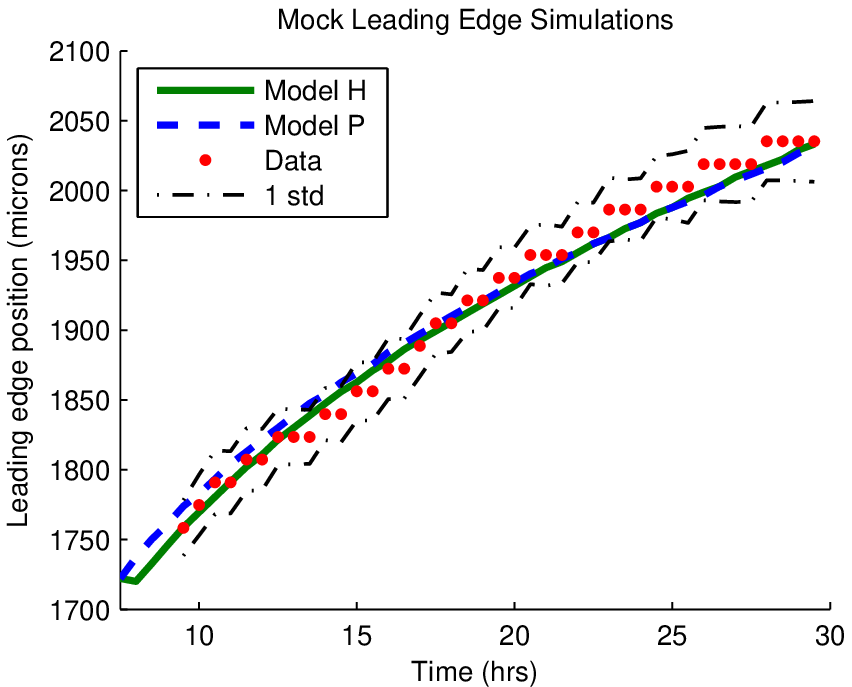}\hfill{}\includegraphics[width=0.45\textwidth]{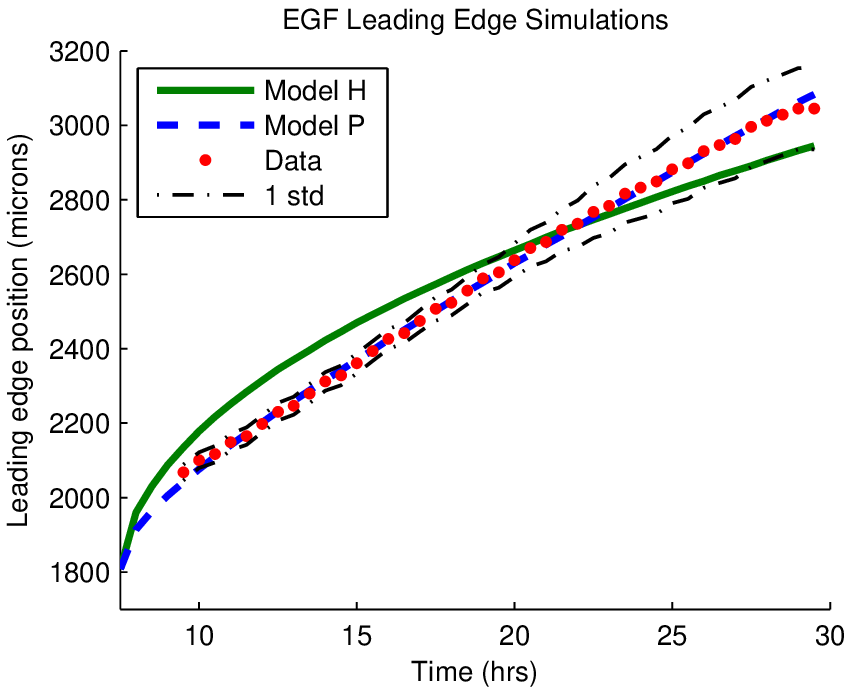}}

\textcolor{black}{\protect\caption{Leading edge simulations for both mock- (left) and EGF- (right) treated
cell populations. The red dots represent the experimental leading
edge locations over time, and the solid green and dashed blue lines
represent the leading edge locations for Models H and P, respectively.
Note that while both models can match the mock data relatively well,
model P appears to fit EGF data better than model H.\textcolor{blue}{{}
}\textcolor{black}{The black dash-dot lines denote one standard deviation
of the leading edge locations, and their computation is described
in the appendix.} \label{fig:Leading-edge-simulations}}
}
\end{figure}

\textcolor{black}{Using the protocol described in Section \ref{sub:Parameter-estimation},
we fit both models to both mock and EGF data. The resulting leading
edge simulations are depicted in Figure \ref{fig:Leading-edge-simulations},
and the parameters used to obtain these plots are given in Table \ref{tab:List-of-parameters.}.
The red dots in Figure \ref{fig:Leading-edge-simulations} represent
the location of the experimental leading edge over time, while the
green and blue lines represent the leading edge locations for Models
H and P, respectively. The thin black dash-dot lines represent one
standard deviation of the data, and we describe their computation
in the appendix. Note that the scale for the two images is very different,
as the leading edge in EGF simulations moves much faster and farther
than the leading edge in mock simulations. Recall that mock leading
edge data sometimes takes the same value for several consecutive time
points because it moves slowly, and the camera discretizes the experimental
domain.} 

\textcolor{black}{In the mock simulations (depicted in the left image
of Figure \ref{fig:Leading-edge-simulations}), both models appear
to match the leading edge data in a similar manner. In particular,
after $t=18$ hours, the two leading edge simulations appear superimposed,
suggesting that these two very different models can yield similar
results for predicting the the leading edge location. Table \ref{tab:List-of-parameters.}
summarizes, however, that these two models use very different simulations
to obtain these similar results. Note that Model H relies primarily
on diffusion for movement, as it uses a large baseline diffusion constant
of $D=19,770\ \mbox{microns}^{2}/\mbox{hr}$, compared to Model P,
which uses a much smaller value of $D=2.19\ \mbox{microns}^{2}/\mbox{hr}$.
On the other hand, Model P uses a large rate of cell-cell adhesion,
as it estimates $\mbox{\ensuremath{\gamma}\ = }30,614\ \mbox{microns}^{2}/\mbox{hr}$
while Model H estimates $\mbox{\ensuremath{\gamma}\ = }1.68\ \mbox{microns}^{2}/\mbox{hr}$
for its rate of cell-cell adhesion. Thus we see that Model H uses
a high rate of diffusion to match the mock data, whereas Model P uses
a high rate of cell-cell adhesion to match this data. These results
are not surprising based on the derivations of these two models, as
cell-cell adhesions hinder movement for Model H but promote movement
for Model P.}

\textcolor{black}{Extending the same parameter estimation methodology
to EGF data, we see on the right hand side of Figure \ref{fig:Leading-edge-simulations}
that Model P fits the experimental data better than Model H. Model
P matches the experimental data well at the initial time point of
$t=10$ hours and continues to do so until the final time point. On
the other hand, Model H initially overestimates the leading edge from
$t=10-20$ hours (so much so that it even fails to match the data
to within one standard deviation at times) and then underestimates
the leading edge location from $t=20-30$ hours. We thus observe that
Model H is unable to accurately simulate the EGF data, suggesting
that MAPK activation may not decrease drag strength during wound healing
when the cell population is treated with EGF.}

\textcolor{black}{Looking at the parameter values for Model P simulations
in Table \ref{tab:List-of-parameters.}, we see that the estimate
for $\gamma$ increases significantly in response to EGF treatment.
More surprising, however, is the decrease in the rate of baseline
cell diffusion that we observe between the mock and EGF simulations
for Model P, from $D=2.19\ \mbox{microns}^{2}/\mbox{hr}$ in the mock
experiment to $D=0.14\ \mbox{microns}^{2}/\mbox{hr}$ in the EGF experiment.
Due to the relative insensitivity of Model P to the parameter D (results
not shown), we can likely consider these values of D as equivalent.
Hence, MAPK activation appears to have a negligible effect on the
baseline rate of diffusion. This is an interesting observation, as
one may initially suspect that increases in MAPK activation increase
the baseline rate of diffusion as a means to stimulate migration,
as we observe for Model H, whose estimates of $D$ change from 19,770
to 177,940 $\text{microns}/\text{h\ensuremath{r^{2}}}$ in response
to EGF treatment.}

\textcolor{black}{}

\subsection{\textcolor{black}{Wave speed comparisons\label{sub:Wavespeed-comparisons}}}

\textcolor{black}{A previous study on predicting the time for wound
closure in intestinal epithelial cells used instantaneous velocity
of the wound edge to compare model simulations with experimental data
\citep{Arciero2013}. In a similar fashion, we compare wave speed
simulations of both the experimental data and the best-fit model simulations
from the previous section. We use a forward difference calculation
on the leading edge locations over time to estimate the model simulation
wave speeds, $w_{m,\beta}^{n}$, for various normalized cellular density
levels (given by $\beta$ = 0.2, 0.3, 0.4):}

\begin{figure}
\subfloat[\textcolor{black}{Wave speed comparisons for EGF scenarios for Model
H (solid line), Model P (dashed line), and experimental data (various
shapes). The wavespeeds for $\beta=0.2$ are compared in green, for
$\beta=0.3$ are in blue, and for $\beta=0.4$ are in red. Note that
model P tends to match the data better than Model H, and that Model
H consistently over and under}estimates almost all wave speed values
from the data for $\beta=0.2$ and $\beta=0.4,$ respectively. \label{fig:Wavespeed-simulations-for}]{\protect\includegraphics[width=0.33\textwidth]{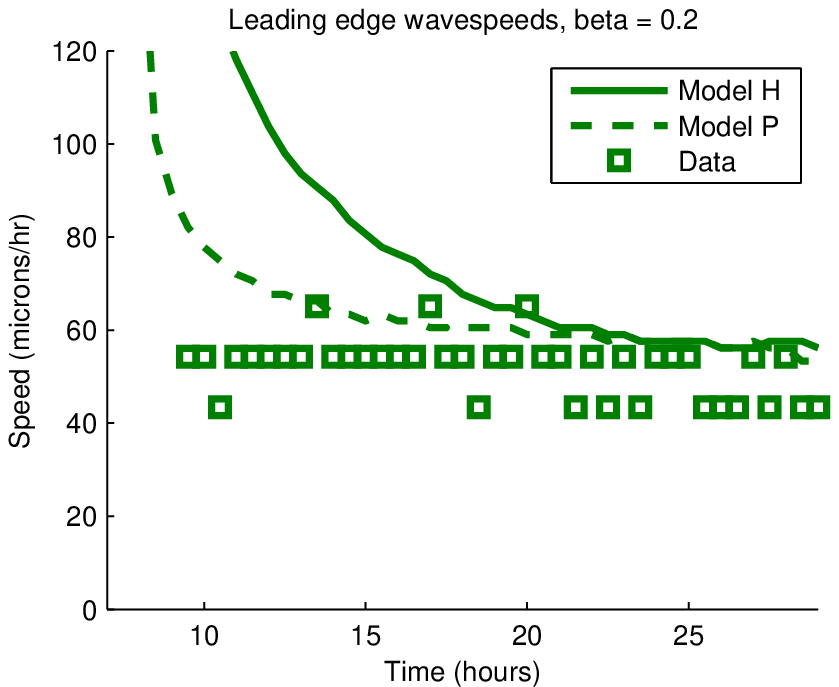}\hfill{}\protect\includegraphics[width=0.33\textwidth]{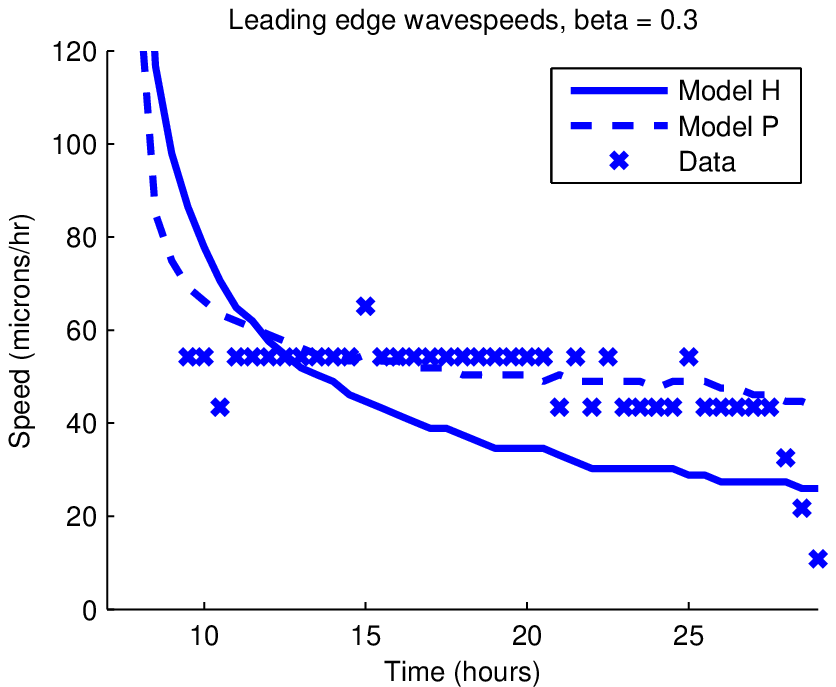}\hfill{}

\protect\includegraphics[width=0.33\textwidth]{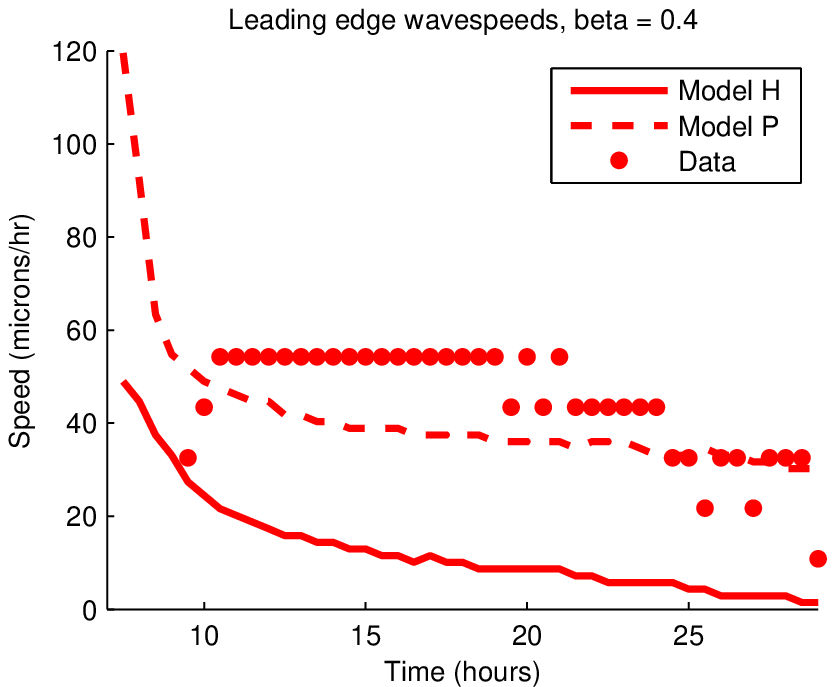}}

\subfloat[\textcolor{black}{Superposition of all density wave speeds with data.
Model H is depicted on the left hand side, and Model P is depicted
on the right hand side. Note that Model P agrees more with the data
than Model H. \label{fig:All-wavespeed-values_superimpose}}]{\centering\protect\includegraphics[width=0.45\textwidth]{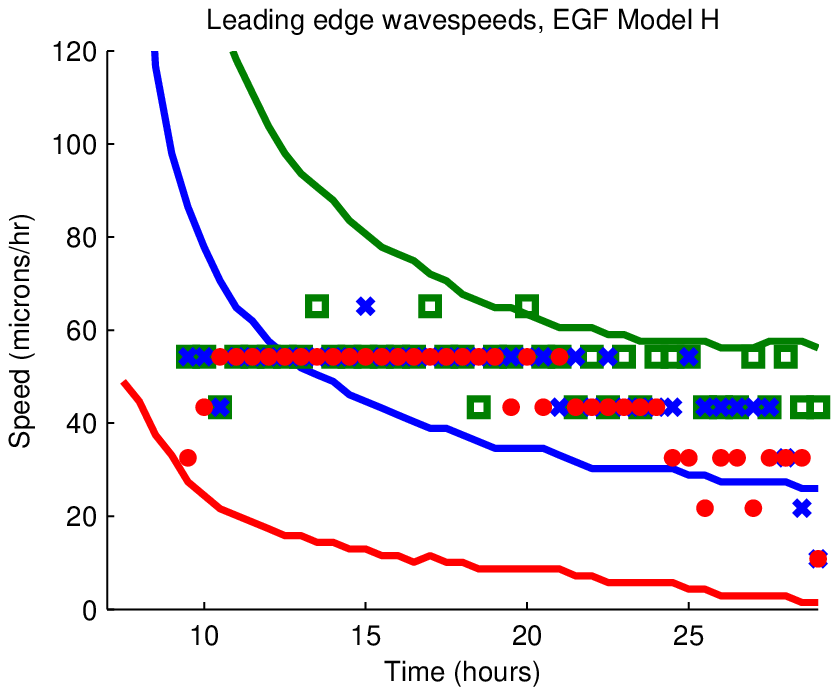}\hfill{}\protect\includegraphics[width=0.45\textwidth]{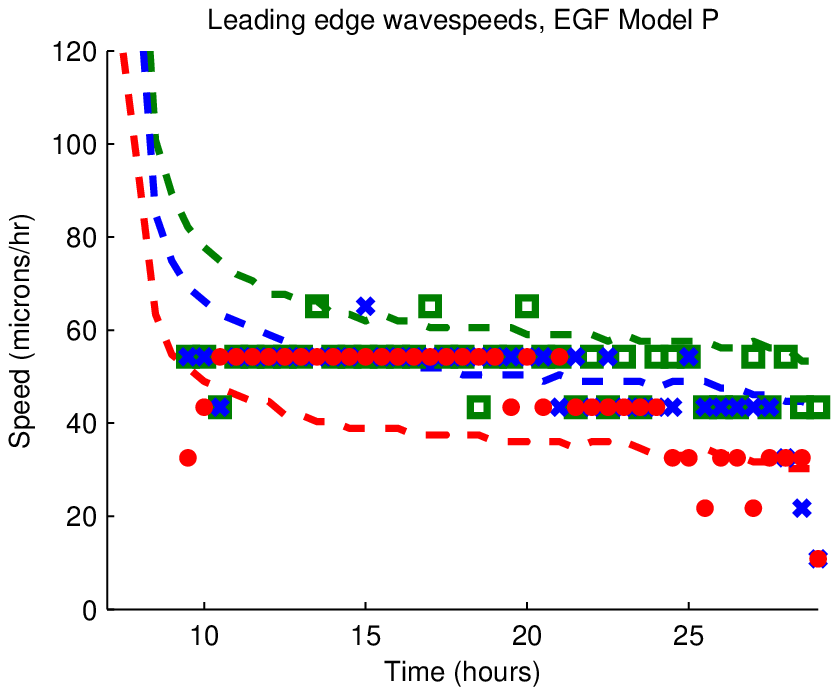}}

\protect\caption{\textcolor{black}{Wave speed simulations for the two models in comparison
with data. }}

\end{figure}

\textcolor{black}{
\[
w_{m,\beta}^{n}=\dfrac{\tilde{\ell}_{m,\beta}^{n+1}-\tilde{\ell}_{m,\beta}^{n}}{\Delta t},
\]
and the data wave speeds are calculated analogously. In this equation,
$\tilde{\ell}_{m,\beta}^{n}$ denotes the 4-point moving average (computed
using MATLAB's smooth function) of the leading edge location at time
$t_{n}$ using $\beta$ as the leading edge threshold. We use $\tilde{\ell}_{m,\beta}^{n}$
instead of $\hat{\ell}_{m,\beta}^{n}$ in this calculation to obtain
realistic wave speed values. Note that the data has a very sharp front,
which causes similar wave speed values for all densities considered
from $t=10-20$ hours, which is demonstrated in Figure \ref{fig:All-wavespeed-values_superimpose}.}

\textcolor{black}{The results for EGF scenario model and data wave
speeds for $\beta=0.2,0.3,0.4$ are depicted in Figure \ref{fig:Wavespeed-simulations-for}
in green, blue, and red, respectively. Note that both model wave speeds
appear to have two different behavior phases: an initial fast phase
(likely due to the presence of empty space), followed by a sustained,
slowly decreasing phase. The previous study also observed these two
phases of behavior and sometimes also observed an increase in speed
towards the end of the simulation due to wound closure. Compare these
changes in phase behavior with that of the the data, which initially
has a constant velocity but begins to slow down after $t=20$ hours.
The initial high velocity of the model simulations (and lack thereof
in the experimental data) is why we let the model simulations run
for 2.5 hours before comparing their leading edges with the experiments.}

\textcolor{black}{In comparing the two models' wave speed simulations
to the data, we see that Model H consistently underestimates the wave
speed for $\beta=0.3,0.4$, and overestimates the wave speed for $\beta=0.2$.
On the other hand, Model P can match the faster phase of the data
for $\beta=0.2$ (around 55 $\mbox{microns}/\mbox{hr}$) from $t=17-30$
hours, and the slower phases of the data (between $35\ \mbox{and }45\ \mbox{microns}/\mbox{hr})$
for $\beta=0.3$ and $\beta=0.4$ from $t=20-29$ hours. Overall,
Model P tends to match the data better than Model H for all time points.
In Figure \ref{fig:All-wavespeed-values_superimpose}, we superimpose
all data and model wave speed values to obtain a better picture of
the overall population wave speeds. Here we again observe more agreement
between the experimental data and Model P than we observe for Model
H. From these simulations, it is not surprising that Model P can more
accurately fit the leading edge data than Model H, as its best-fit
simulation better matches the wave speeds of the data. }

\subsection{\textcolor{black}{Comparison to data snap shots\label{sub:Comparison-to-data}}}

\begin{figure}
\textcolor{black}{\includegraphics[width=0.45\textwidth]{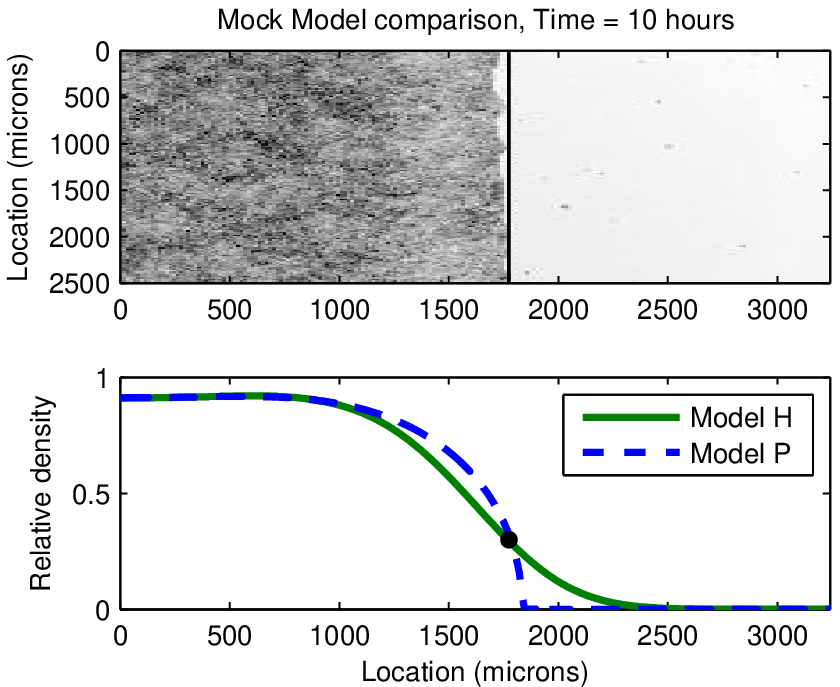}\hfill{}\includegraphics[width=0.45\textwidth]{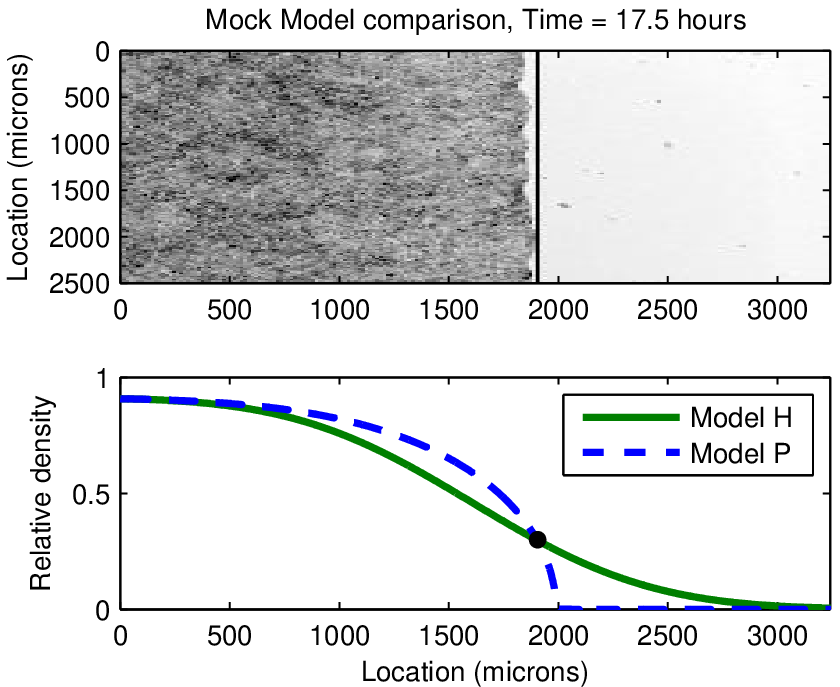}}

\textcolor{black}{\includegraphics[width=0.45\textwidth]{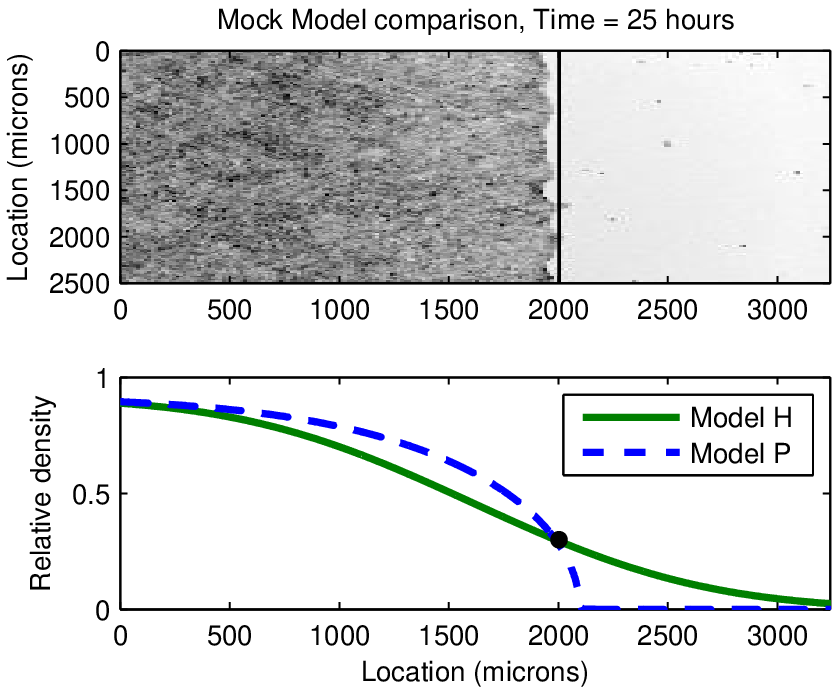}\hfill{}\includegraphics[width=0.45\textwidth]{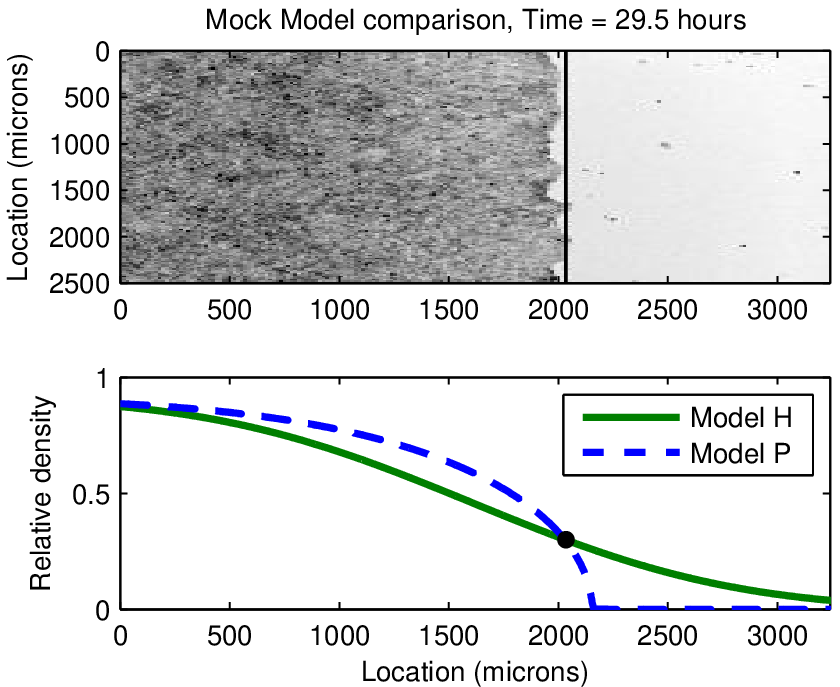}\protect\caption{Comparisons of experimental images (upper frames) to model profiles
(lower frames) for the mock experiments over various time snapshots.
\textcolor{black}{The black line in the upper frames and dot in the
bottom frames depict the experimental leading edge over time. }Note
the gradually decreasing cell front of Model H in comparison to the
sharp front of Model P. \label{fig:mock_exp}}
}
\end{figure}

\begin{figure}
\textcolor{black}{\includegraphics[width=0.45\textwidth]{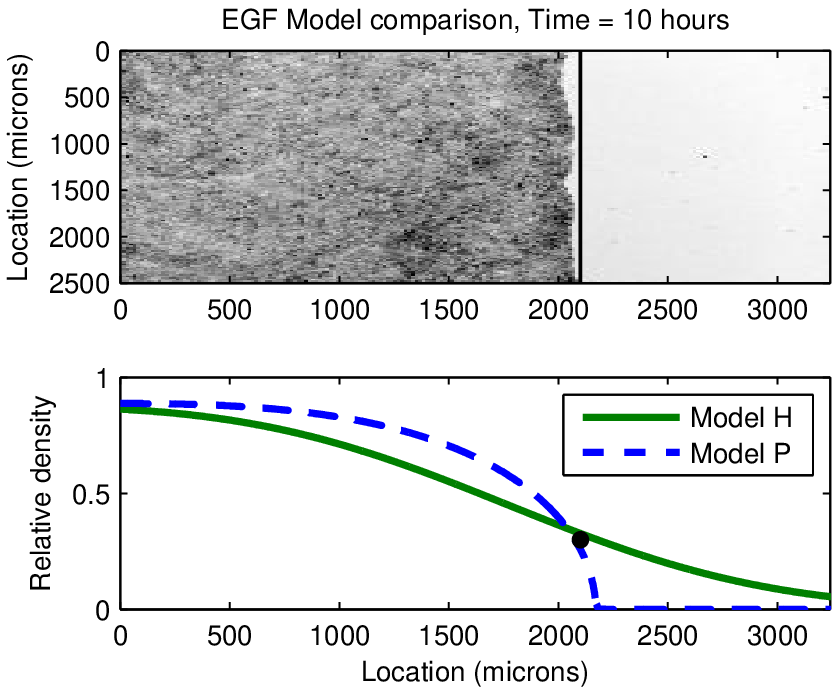}\hfill{}\includegraphics[width=0.45\textwidth]{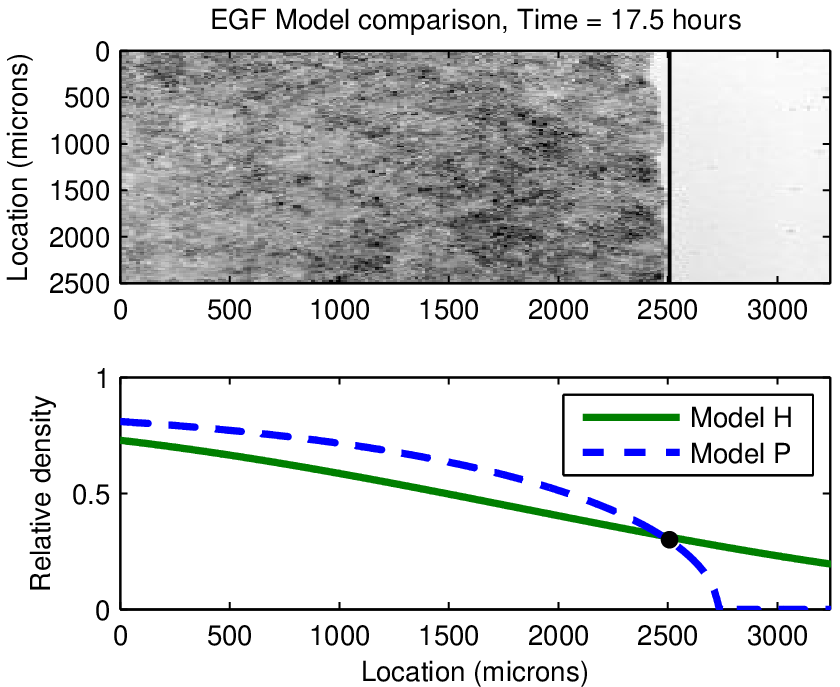}}

\textcolor{black}{\includegraphics[width=0.45\textwidth]{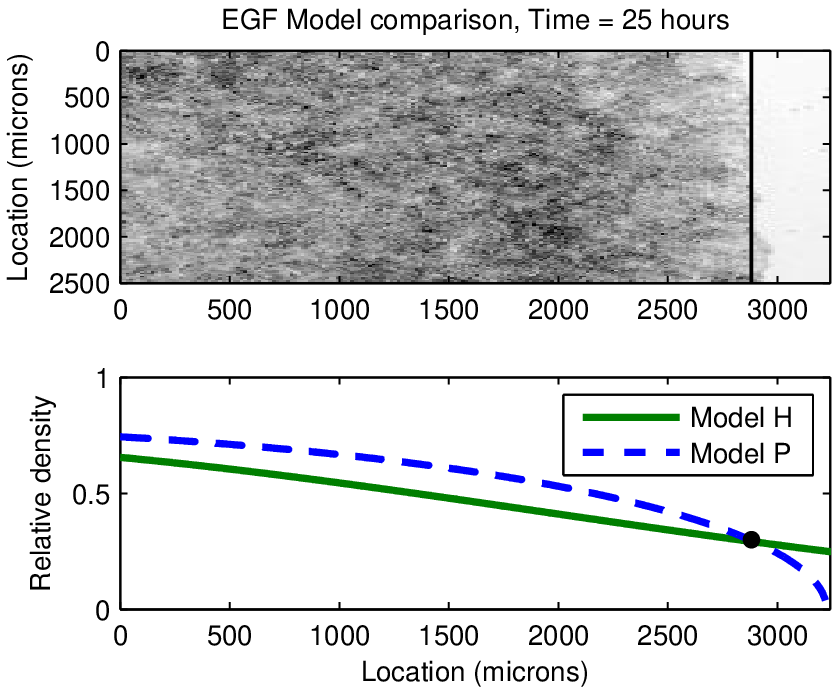}\hfill{}\includegraphics[width=0.45\textwidth]{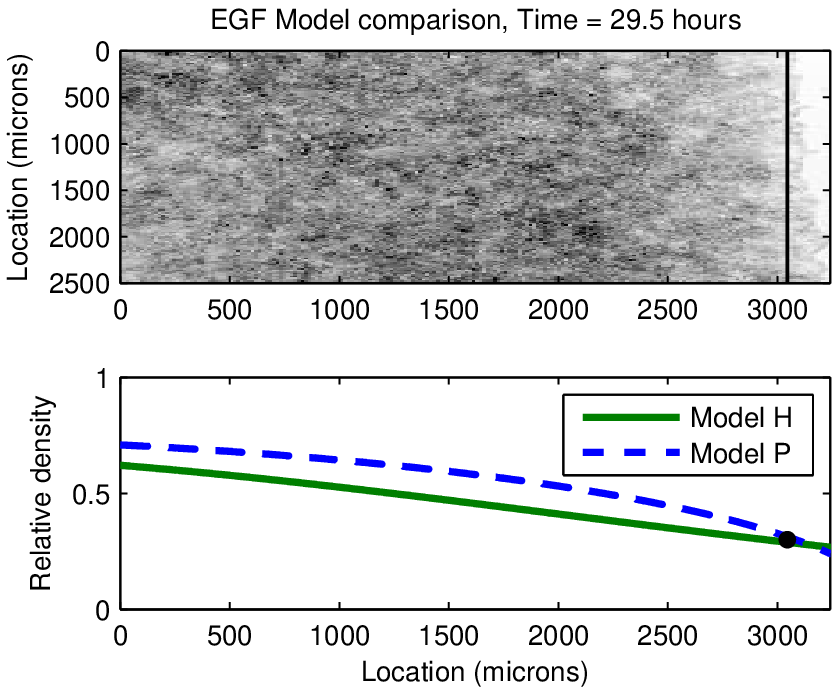}\protect\caption{Comparisons of experimental images (upper frames) to model profiles
(lower frames) for the EGF experiments over various time snapshots.
\textcolor{black}{The black line in the upper frames and dot in the
bottom frames depict the experimental leading edge over time. }Note
the gradually decreasing cell front of Model H in comparison to the
sharp front of Model P. Observe how this gradual population frony
causes Model H to predict a nonzero value for \textcolor{black}{$u(t=17.5,x=3000)$,}
even though there are no cells present at this time and location in
the experiment. \label{figLegf_exp}}
}
\end{figure}

\textcolor{black}{We next compare the model profiles to experimental
snapshots over time for both mock and EGF simulations. These images
are depicted in Figures \ref{fig:mock_exp} and \ref{figLegf_exp}
for mock and EGF data, respectively, for $t=10,17.5,25,\mbox{ and }29.5$
hours. In each snapshot, the top frame represents the experimental
image of the cell sheet during the wound healing experiment, and the
bottom frame demonstrates the model simulations for both Models H
and P. In each snapshot, the experimental leading edge is depicted
with a black line in the top frame and a black dot in the bottom frame.}

\textcolor{black}{In Figure \ref{fig:mock_exp}, we note that both
Models H and P match the experimental leading edge well, as we would
expect from Figure \ref{fig:Leading-edge-simulations}. However, the
two resulting profiles are very different, as Model H has a gradually
decreasing profile in comparison to the steep population front of
Model P. Considering our previous observation that Model H uses a
high rate of diffusion to match the leading edge, whereas Model H
uses a high rate of cell-cell adhesion, these two different model
profiles are not surprising. Note that this gradually descending cell
profile causes Model H to predict a significant presence of cells
in locations that are clearly empty in the experiment. For example,
at $t=29.5$ hours, Model H predicts a cellular density of about $u(t=29.5,x=2500)$
= 0.25 (one quarter of the maximum cellular density at $x=$ 2500
microns), whereas the experimental cell sheet has clearly not yet
reached this location, and its leading edge is approximately 500 microns
further back. We believe this difference in profile front descent
allows Model P to better predict the leading edge location and various
wave speeds of the experimental wound healing data.}

\textcolor{black}{}

\textcolor{black}{Figure \ref{figLegf_exp} depicts the same figure
for the EGF experiments and simulations. We again notice that both
profiles tend to agree near the leading edge location, but yield very
different profile simulations. Model H's profile again descends much
more gradually than the steep front of Model P. Similarly, Model H
also predicts the presence of cells in areas that are clearly empty
in the experimental snapshot, such as at $x=3000$ microns at $t=17.5$
hours. We note, however, that Model P also predicts the presence of
cells in clearly empty areas from the EGF experiment (such as past
$x=3000$ microns at $t=25$ hours). This suggests that more sophisticated
modeling endeavors are still needed in the future to better capture
the leading edge dynamics and match the experimental profiles of wound
healing data. Overall, however, Model P can accurately match the leading
edge dynamics well for both mock and EGF simulations, whereas Model
H is considerably flawed in several aspects. These results support
our second hypothesis that }\textcolor{black}{\emph{MAPK activation
in response to EGF treatment stimulates collective migration by increasing
the pulling strength of leader cells.}}

\textcolor{black}{}

\subsection{\textcolor{black}{Time to wound closure\label{sub:Time-to-wound}}}

\begin{figure}
\centering{}\includegraphics[width=0.45\textwidth]{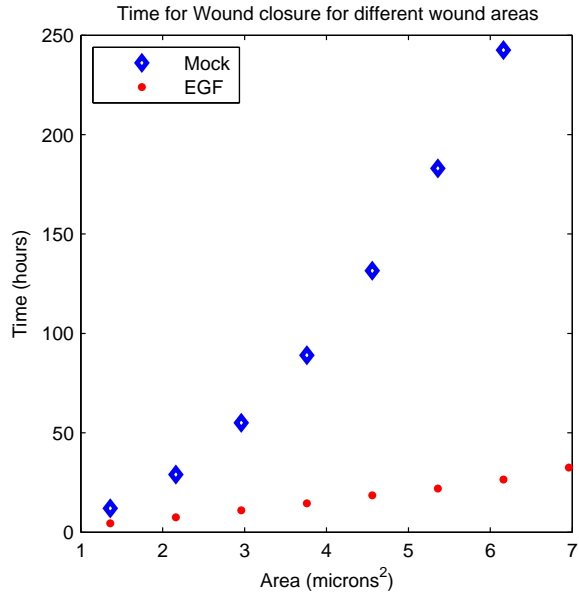}\protect\caption{Using Model P to predict the time to wound closure as a function of
wound area. The blue diamonds denote closure time for mock cells,
whereas the red dots denote closure time for EGF-treated cells. Note
that the time to wound closure for mock-treated cells appears quadratic,
which matches results from \citep{Arciero2013}. The time to wound
closure appears to become linear in response to EGF treatment. \label{fig:Time-of-wound_closure}}
\end{figure}

\textcolor{black}{A common theme in wound healing studies is to investigate
how different wound aspects, including size, shape, and aspect ratio,
affect the time to wound closure \citep{Arciero2013,Gilman2004}.
Times for wound closure as a function of wound area were analyzed
in \citep{Arciero2013} for different wound shapes and sizes. In a
similar fashion, we can use Model P to predict the closure times for
various different wound sizes, as is depicted in Figure \ref{fig:Time-of-wound_closure}.
Because this model is simulated in one dimension, we only use it on
rectangular wound shapes, where we calculate the wound area by multiplying
the length of the wound by the width of the well (4 microns). We also
define the time to wound closure as the time it takes for the leading
edge to reach the end of the field of view (3,240 microns), or $t_{k}$
such that $\hat{\ell}_{m,0.3}^{k}=3,240$ microns. Notice in this
figure how the time to wound closure for mock cells appears approximately
quadratic, which would be consistent with \citep{Arciero2013}, in
which the authors suggested that the rate of change of wound area
is proportional to the area of the wound in IEC cells. In response
to EGF treatment, however, we see this curve change to a much faster
and apparently linear term. We can validate these predicted times
in future studies by experimenting with wounds of various sizes. We
may also investigate nonrectangular wounds in the future by extending
our model to two dimensions.}

\begin{doublespace}

\section{\textcolor{black}{Conclusions and Discussion\label{sec:Discussion}}}
\end{doublespace}

\textcolor{black}{In this study, we have developed two mathematical
models representing the spread of a cellular sheet during wound healing
in response to EGF treatment. These two models are used to test two
hypotheses on the effects of MAPK activity on cell-cell adhesion during
collective migration. Model H assumes that MAPK activity stimulates
collective migration through decreases in the drag strength of follower
cells, whereas Model P assumes that MAPK activity stimulates collective
migration through increases in the pulling strength of leader cells.
Model P matches several aspects of the experimental data better than
Model H, ultimately suggesting the validity of our second hypothesis.
From our resulting parameter values, we also observe that the main
effect of MAPK activity on cell migration is an increase in the rate
of cell-cell adhesion, and that it has a negligible effect on the
baseline rate of diffusion. As one may initially suspect that increases
in MAPK activation stimulate the baseline rate of diffusion to promote
the rate of migration, we plan to further investigate this observation
in future experimental and computational studies due to its nonintuitive
nature}

\textcolor{black}{These conclusions have implications in our understanding
of the EMT, in which cells detach from an epithelial cell population
as mesenchymal cells \citep{Chapnick2011,Friedl2009}. Cells undergoing
EMT tend to gain migratory and invasive properties, and hence the
EMT is a driving factor in both normal embryological development and
cancer metastasis \citep{Chapnick2011,Janda2002}. There are two main
types of EMT: incomplete and complete. During incomplete EMT, cell-cell
junctions remain intact between certain invasive cells, and collective
strand-like migration is observed. Complete EMT is characterized by
the loss of most cell connections, and individual cells detach from
the population to begin their own migration pattern. Hence, in either
form of EMT, we see that cells detach from the population and can
be considered smaller cell subpopulations (Composed of one or several
cells), effectively increasing the prevalence of leader cells. Since
cell pulling appears to be the dominant mechanism of collective migration,
we suggest that HaCaT cells in response to EGF treatment may express
incomplete EMT behaviors so that an increased number of leader cells
can effectively guide smaller cell subpopulations via pulling. We
may use this observation to better understand incomplete EMT dynamics
through the experimental system and mathematical analysis, along with
an understanding of any possible connections between incomplete and
complete EMT. For example, if any sort of chemical treatment or oncogenic
perturbations will convert the system from incomplete to complete
EMT or vice versa.}

\textcolor{black}{}
\begin{figure}
\centering{}\textcolor{black}{\includegraphics[width=0.59\textwidth,height=8cm]{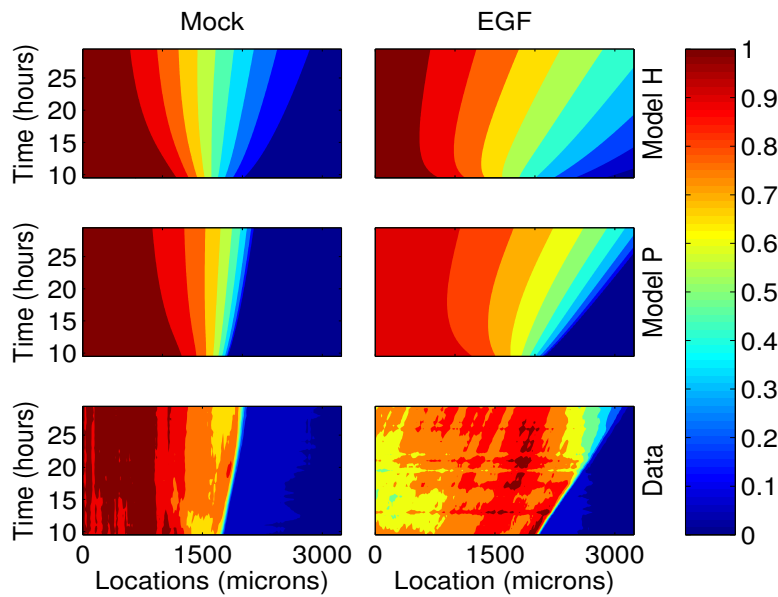}\protect\caption{Contours for the two models (top two rows) against data (bottom row).
The $x-$axis depicts the spatial location and the $y-$axis depicts
time to demonstrate how the profiles change over time for both model
simulations and the actual experiment. The colorbar depicts the cellular
density levels (i.e., blue denotes a low density and red denotes a
high density). \label{fig:Contours-for-the}}
}
\end{figure}

\begin{doublespace}
\textcolor{black}{In Figure \ref{fig:Contours-for-the}, we have depicted
several data and model contours over time. From these contours, we
see that Model P is able to accurately match the leading edge data
for both mock and EGF data, but it is still far from matching the
entire data profile accurately over time. Future endeavors will aim
to match the entire population profile instead of just the leading
edge. We note that the role that the MAPK signaling cascade on cell-surface
adhesion during collective migration is another important question
that requires investigation. Further exploration into this may aid
in better matching the entire cell population profile. We also note
that we used a simple linear term for average MAPK activation levels
in our model. The slope of this line very likely depends on the amount
of EGF treatment in the population, however, so future studies will
include more sophisticated modeling terms that may involve EGF ligand,
its intracellular precursor, and reactive oxygen species, as has been
done in a previous model on MAPK activation during wound healing\citep{Posta2010}.
EGF ligand secretion by leader cells likely aids in collective migration
as ligand diffusion to follower cells will further promoting cell
pulling in the population.}
\end{doublespace}

\textcolor{black}{In this work, we have developed a simple model to
investigate how EGF treatment influences collective migration during
wound healing in HaCaT cells. We demonstrate that cell-cell adhesion
plays a critical and activating role in collective migration through
leader cell pulling. This model may have implications in understanding
EMT dynamics, including how cells may transition from incomplete to
complete EMT and vice versa.}

\begin{doublespace}

\section*{\textcolor{black}{Acknowledgements}}
\end{doublespace}

\begin{doublespace}
\textcolor{black}{JTN is supported by the Interdisciplinary Quantitative
Biology (IQ Biology) program at the BioFrontiers Institute, University
of Colorado, Boulder, which is supported by NSF IGERT grant number
1144807. This work was in part supported by grants from National Institutes
of Health R01CA107098 to X.L. The authors would like to thank Joe
Dragavon and the BioFrontiers Advanced Light Microscopy Core for their
microscopy support. The ImageXpress MicroXL was supported by a NCRR
grant S10 RR026680 from the NIH. }
\end{doublespace}

\section*{\textcolor{black}{Appendix}}

\subsection*{\textcolor{black}{Numerical implementation}}

\begin{doublespace}
\textcolor{black}{Note that for numerical implementation, we use a
method of lines approach with MATLAB's ode15s function to simulate
the entire model. For spatial discretization, we use the second order
scheme for convection-diffusion equations (without convection in our
case) from \citep{Kurganov2000} given by 
\[
\dot{u}_{j}(t)=\dfrac{P_{j+\nicefrac{1}{2}}(t)-P_{j-\nicefrac{1}{2}}(t)}{\Delta x}
\]
where $P_{j+\nicefrac{1}{2}}(t)$ is an approximation to the diffusive
flux, given by 
\[
P_{j+\nicefrac{1}{2}}(t)=\dfrac{1}{2}\left[Q\left(u_{j}(t),\dfrac{u_{j+1}(t)-u_{j}(t)}{\Delta x}\right)+Q\left(u_{j+1}(t),\dfrac{u_{j+1}(t)-u_{j}(t)}{\Delta x}\right)\right]
\]
where $Q(u,u_{x})$ denotes the cellular diffusion rate. Simulations
not shown demonstrate quick convergence in our simulations.}
\end{doublespace}

\subsection*{\textcolor{black}{Computing standard deviations}}

\begin{figure}
\textcolor{black}{\includegraphics[width=0.45\textwidth]{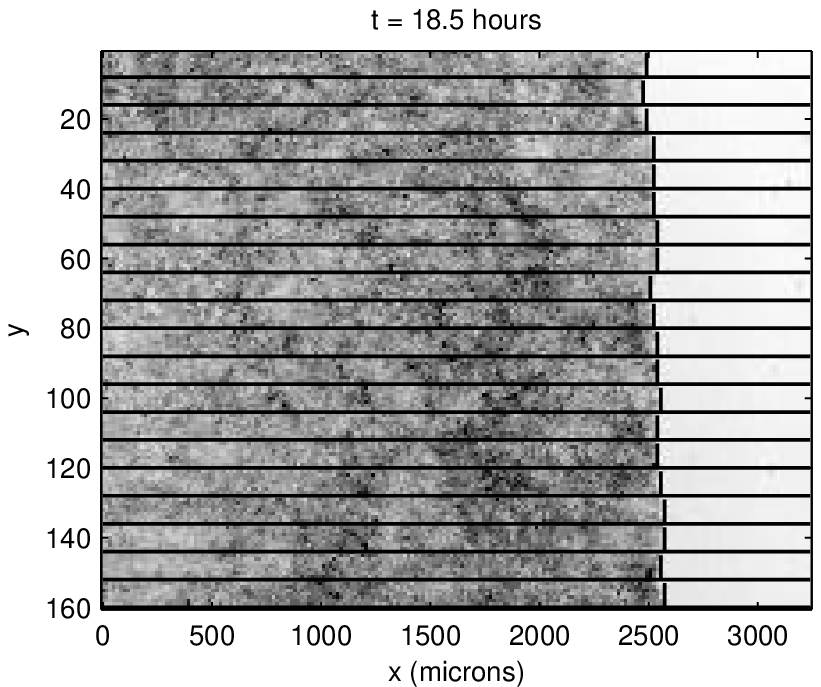}\hfill{}\includegraphics[width=0.45\textwidth]{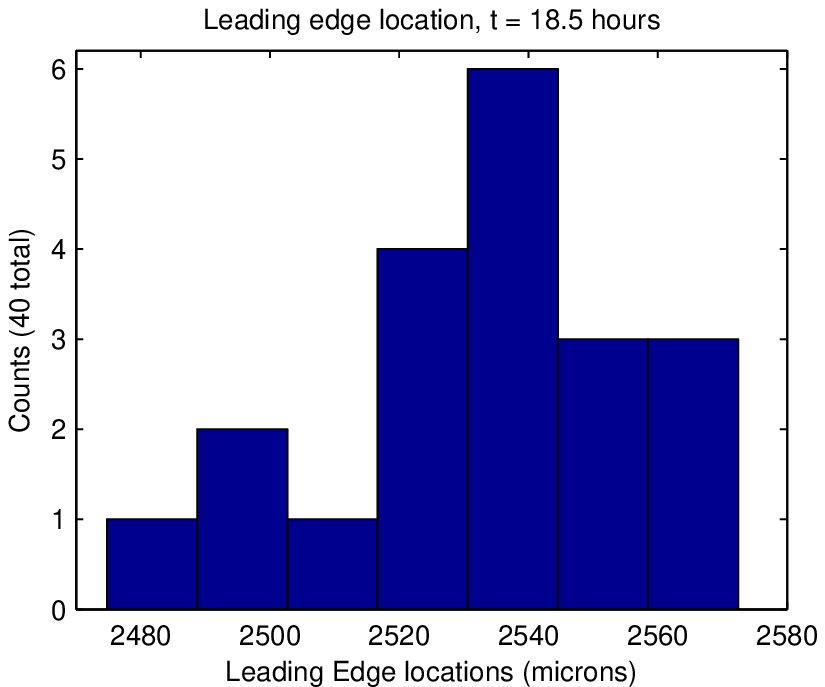}}

\textcolor{black}{\protect\caption{\textcolor{black}{Splitting up the y-axis into twenty bins and obtaining
several estimates for the leading edge. On the left hand image, we've
demonstrated how we split up the axis to calculate several different
leading edge locations (denoted by black vertical lines). On the right
hand image, we show the histogram of the leading edge locations of
the left hand image \label{fig:LE_binning}. }}
}
\end{figure}

\textcolor{black}{In order to compute the standard deviation of the
location of the leading edge at each time point, we split the $y-$axis
of each image into $R$ bins, and compute the leading edge for each
of the separate bins to obtain $R$ estimates of the leading edge
location in the population. In Figure \ref{fig:LE_binning}, for example,
we've split the $y-$axis into twenty separate bins on the right hand
side and calculated the leading edge of the data for each one. Hence,
for each time point during the experiment, we compute the sample mean
of the leading edge locations (call it $\hat{\mu}$) then compute
the sample standard deviation to obtain one standard deviation. On
the right side of Figure \ref{fig:LE_binning}, we've also depicted
an example histogram for the leading edge locations when using twenty
bins.}

\textcolor{black}{}

\textcolor{black}{\bibliographystyle{apalike}
\phantomsection\addcontentsline{toc}{section}{\refname}\bibliography{mathbioCU}
}
\end{document}